\documentclass[apj,usenatbib, iop, twocolappendix]{emulateapj}

\usepackage{comment}
\usepackage{ulem}
\usepackage{tikz}
\usepackage{color}
\usepackage{graphicx}
\usepackage{amsmath}
\usepackage{amsfonts}
\usepackage{outlines}
\usepackage{enumitem}
\usepackage{natbib}
\usepackage{booktabs}

\usepackage[breaklinks,colorlinks,
   urlcolor=blue,citecolor=blue,linkcolor=blue]{hyperref}
\usepackage[all]{hypcap}

\newcommand{\Lrad}{L_{\rm rad}}
\newcommand{\Mdot}{\dot{M}_{\rm w}}
\newcommand{\Edot}{\dot{E}_{\rm w}}

\newcommand{\Ledd}{L_{\rm Edd}}

\newcommand{\beq}{\begin{equation}}
\newcommand{\eeq}{\end{equation}}
\newcommand{\bea}{\begin{eqnarray}}
\newcommand{\eea}{\end{eqnarray}}

\newcommand{\trm}[1]{\textrm{#1}}

\newcommand{\diff}{\text{d}}

\shorttitle{Winds from X-ray Bursts}
\shortauthors{Yu \& Weinberg}

\begin{document}

\defcitealias{Weinberg:06}{WBS}
\defcitealias{intZand:10}{iZW10}


\title{Super-Eddington winds from Type I X-ray bursts}


\author{Hang Yu$^{1}$ and Nevin N. Weinberg$^{1}$}
\affil{$^{1}$Department of Physics, and Kavli Institute for Astrophysics and Space Research, Massachusetts Institute of Technology, \\Cambridge, MA 02139, USA} 
\email{hyu45@mit.edu}
\email{nevin@mit.edu}




\begin{abstract}
We present hydrodynamic simulations of spherically symmetric super-Eddington winds from radius-expansion type I X-ray bursts. Previous studies assumed a steady-state wind and treated the mass-loss rate  as a free parameter.  Using MESA, we follow the multi-zone time-dependent burning, the convective and radiative heating of the atmosphere during the burst rise, and the launch and evolution of the optically thick radiation-driven wind as the photosphere expands outward to radii $r_{\rm ph} \ga 100\trm{ km}$.  We focus on neutron stars (NSs) accreting pure helium and study bursts over a range of ignition depths.  We find that the wind ejects $\approx 0.2\%$ of the accreted layer, nearly independent of ignition depth. This implies that $\approx 30\%$ of the nuclear energy release is used to unbind matter from the NS surface.   We show that ashes of nuclear burning are ejected in the wind and dominate the wind composition for bursts that ignite at column depths $\ga 10^9\trm{ g cm}^{-2}$.  The ejecta are composed primarily of elements with mass numbers $A> 40$, which we find should imprint photoionization edges on the burst spectra.  Evidence of heavy-element edges has been reported in the spectra of strong, radius-expansion bursts.  We find that after $\approx 1\trm{ s}$ the wind composition transitions from mostly light elements ($^4$He and $^{12}$C), which sit at the top of the atmosphere, to mostly heavy elements ($A>40$), which sit deeper down.  This may explain why the photospheric radii of all superexpansion bursts show a transition after $\approx 1\trm{ s}$ from a superexpansion  ($r_{\rm ph}>10^3\trm{ km}$) to a moderate expansion ($r_{\rm ph}\sim 50\trm{ km}$). 
\end{abstract}





\section{Introduction}

Type I X-ray bursts are powered by unstable thermonuclear burning of  accreted material on the surface of a neutron star (NS) in a low-mass X-ray binary (for reviews, see \citealt{Bildsten:98, Strohmayer:06, Galloway:17}).  The peak luminosity and duration of a burst depends primarily on the accretion rate and  composition of the accreted material.  In  photospheric radius expansion (PRE) bursts, which comprise about 20\% of all bursts \citep{Galloway:08}, the luminosity exceeds the Eddington luminosity and radiation forces drive an optically thick wind that lifts the photosphere off the NS surface.  Typically, the photosphere moves out to radii $r_{\rm ph}\approx \trm{few}\times\left(10-100\right)\trm{km}$, although in a small fraction of PRE bursts, known as superexpansion bursts, $r_{\rm ph} \ga 10^3\trm{ km}$ (\citealt{intZand:10}; hereafter \citetalias{intZand:10}). 
As the emitting area of the photosphere increases, its temperature  decreases below $1\trm{ keV}$, leading to a substantial loss of signal for detectors that lack sensitivity at low X-ray energies.
Depending on the ignition depth and hence the total nuclear energy release, the entire PRE can last from seconds to minutes.

In order to reliably interpret observations of PRE bursts, it is important to understand the dynamics of the wind.  Three recent developments particularly motivate such a study: (i) the renewed effort to use PRE bursts to measure NS radii and thereby constrain the NS equation of state (see, e.g., \citealt{vanParadijs:79, Ozel:10, Ozel:16a, Steiner:10, Steiner:13}), (ii) evidence of heavy-element absorption features in burst spectra, which might be imprints of ejected ashes of nuclear burning  (\citetalias{intZand:10}, \citealt{Barriere:15, Iwai:17, Kajava:17}), and (iii) the sensitivity of the {\it Neutron Star Interior Composition Explorer} ({\it NICER}; \citealt{Gendreau:17}) down to 0.2\trm{ keV}. This makes {\it NICER}  an ideal instrument to study strong PRE bursts at high time resolution since, unlike the Proportional Counter Array (PCA; \citealt{Jahoda:06}) on board the {\it Rossi X-ray Timing Explorer} ({\it RXTE}), it does not lose signal when the temperature decreases during the expansion (see \citet{Keek:18}, who studied the first strong PRE burst detected with {\it NICER}).

Constraining NS radii with PRE bursts relies on measuring the flux when the temperature reaches a maximum. This is thought to be the moment when the photosphere ``touches down"  on the NS surface at the end of the PRE.  By knowing the distance to a source and associating the touchdown flux with the Eddington flux, it is possible to constrain the NS radius $R$.  However, the measurements may be subject to considerable systematic errors  \citep{Boutloukos:10, Steiner:10, Suleimanov:11, Miller:13, Medin:16, Miller:16}.  This is partly due to spectral modeling uncertainties, such as how the color-correction factor, which enters the fit to $R$, depends on luminosity and composition.  Recently, there has been considerable effort to make progress on this front \citep{Suleimanov:11, Suleimanov:12, Medin:16}, including the study by \citet{Nattila:15}, who showed that the emergent spectra are sensitive to the abundance of heavy elements in the wind. There are also uncertainties associated with the dynamics of the PRE.  For example, \citet{Steiner:10} found it necessary to relax the assumption that $r_{\rm ph}=R$ at touchdown in order to avoid unphysical values of NS mass and radius when fitting to PRE burst data.  A better understanding of the dynamics of the PRE and the wind composition could help address these uncertainties.   

\citeauthor{Weinberg:06} (2006; hereafter  \citetalias{Weinberg:06}) modeled the evolution of the atmosphere during the rise of a PRE burst.  However, they only considered times up to when the luminosity first reaches the Eddington luminosity; they did not study the dynamics of the subsequent PRE.  Nonetheless, their calculations suggested that the wind could eject ashes of nuclear burning.  This is because during the burst rise, there is an extensive convective region that is well mixed with ashes brought up from the burning layer below. Based on approximate energetic arguments, they estimated that the wind would be launched from a region that contains ashes and thereby expose them during (and after) the PRE. Since the ashes are primarily heavy elements (mass numbers $A\sim 30-60$; see, e.g., \citetalias{Weinberg:06}), they could imprint absorption edges and lines on the burst spectra. A detection would probe the nuclear burning processes and might enable a measurement of the gravitational redshift of the NS. The latter possibility assumes that the heavy elements do not sink too quickly once the photosphere settles back to the NS surface at the end of the PRE; we will show that for deep ignitions there are very few light elements in the photosphere relative to which the ashes could sink, which suggests that the ashes may indeed linger on the surface.

Two superexpansion bursts detected with {\it RXTE} from 4U 0614+091 and 4U 1722-30 showed significant deviation from an absorbed blackbody \citepalias{intZand:10}. By including an absorption edge in the spectral model, \citetalias{intZand:10} found that they could significantly improve the fits to the data.  The energy of the fitted edges was consistent with the H-like photoionization edge of Ni and the optical depth of the edges suggested Ni mass fractions $X\ga 0.1$.  \citet{Kajava:17} detected similar features in the spectra of an {\it RXTE} burst from HETE J1900.1-2455. In {\it NuSTAR} observations of a burst from GRS 1741.9-2853, \citet{Barriere:15} detected,  at a $1.7\sigma$ confidence level, a narrow absorption line at $5.46\pm 0.10\trm{ keV}$.  They proposed that the line, if real, formed in the wind above the photosphere by a resonant K$\alpha$ transition from H-like Cr.  

Although including heavy-element absorption features improved the fits to these bursts, the limited spectral resolution of the PCA on {\it RXTE} and the weakness of the {\it NuSTAR} spectral line preclude an unambiguous identification.   The only high-spectral-resolution observations of PRE bursts from any source to date are six bursts detected with {\it Chandra} from 4U 1728-34 \citep{Galloway:10}  and one from SAX J1808.4-3658 observed simultaneously with {\it Chandra} and {\it RXTE} \citep{intZand:13}.  No discrete features were detected in the spectra, although this might be because the radius expansions were all weak ($r_{\rm ph}\approx 20\trm{ km}$); the upper limits on the edge equivalent widths were a few hundred eV, comparable with the predictions of \citetalias{Weinberg:06}.
 
All previous models of PRE assumed a steady-state wind (i.e., time-independent models). The first models were Newtonian  \citep{Ebisuzaki:83, Kato:83, Quinn:85, Joss:87} and then fully relativistic \citep{Paczynski:86}. By developing an  improved treatment of radiative transfer, \citet{Joss:87} constructed models that extended into the optically thin regions above the photosphere, where Compton scattering is important.  These models all treated the wind mass-loss rate $\Mdot$ as a free parameter.  \citet{Nobili:94} removed $\Mdot$ as a free parameter by including nuclear energy generation due to helium burning in the innermost regions of their model (their models were also relativistic and improved upon previous treatments of radiative transfer). However, most of the energy released from helium burning occurs within a few milliseconds, well before the wind is launched \citepalias{Weinberg:06}. We will show that in order to properly account for the driving of the wind, it is necessary to consider the transport of heat (by convection and radiative diffusion) through the hydrostatic layers between the ignition base and the wind base.

We find that it takes a few seconds for nearly time-independent conditions to be established in the wind (see also table 1 in \citealt{Joss:87}).  Since most observed PREs only last for a few seconds, the steady-state assumption is often not well satisfied. This, and the recent developments discussed above, motivate a time-dependent calculation of the wind.  

Once He ignites, the calculation can be divided into two time-dependent stages: a hydrostatic heating stage (the burst rise) followed by a hydrodynamic wind stage (the PRE phase).  In the first stage, which we study in Section~\ref{sec:num_mlt}, the atmosphere above the helium burning layer is heated by convection and radiative diffusion.  Initially, the radiative heat flux is sub-Eddington and the atmosphere adjusts hydrostatically.  During this time, freshly synthesized ashes are dredged up by convection and mixed throughout the growing  convective region.  As the atmosphere heats up, the radiative flux increases and eventually exceeds the local Eddington limit at the top of the atmosphere.  This marks the beginning of the second stage, the PRE, which we study in  Section~\ref{sec:num_wind}. We show that as the photosphere expands outward, the base of the wind moves downwards to greater depths. First it blows away the top most layers of the atmosphere, which consists mostly of light elements.  But gradually it digs into the deeper layers and ejects heavy-element ashes.  In Section~\ref{sec:observational_signatures} we describe the observational signatures of the wind models and compare them with observed PRE bursts. Although our results are broadly consistent with observations, there are also some notable differences.  We consider whether these might be attributed to some of our simplifying assumptions, including our neglect of general relativistic effects and our simplified treatment of radiative transfer, which relies on the diffusion approximation and neglects potential line-driving of the heavy elements.  In Section~\ref{sec:summary} we summarize and conclude.

\section{Hydrostatic Burst Rise}
\label{sec:num_mlt}

\begin{figure}
\includegraphics[trim=-1.5cm 0 0 0.5cm, width=0.47\textwidth]{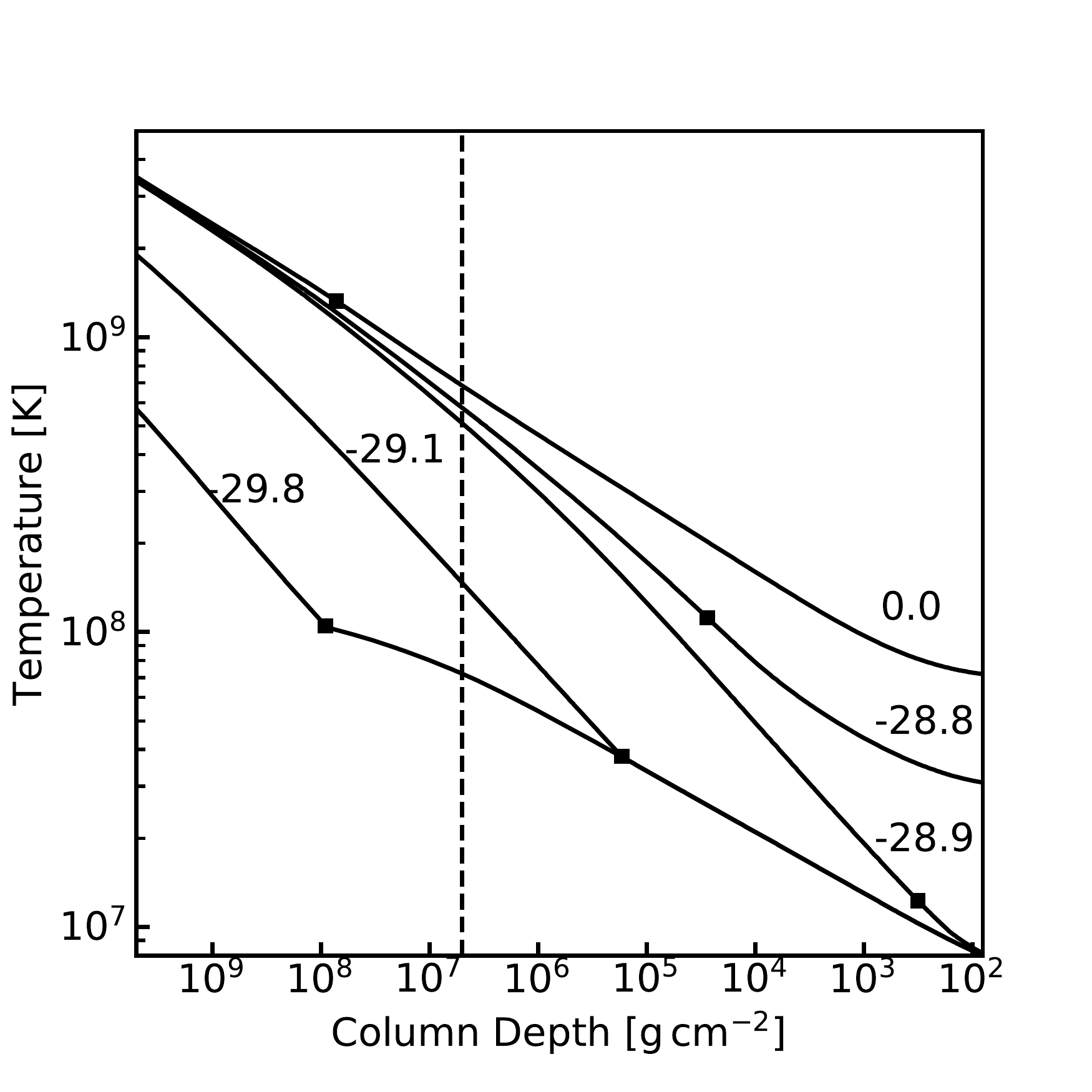}
\caption{Temperature as a function of column depth for model y3n21  at different moments during the burst rise. The numbers label the time in milliseconds, with $t=0.0$ corresponding to when $\Lrad$ first exceeds $\Ledd$.  The squares indicate the top of the convective zone and the dashed vertical line indicates the maximum depth of the wind base $y_{\rm wb}$ during the PRE phase; material above this line will  be ejected by the wind.}
\label{fig:convZone}
\end{figure}

We model the hydrostatic portion of the burst rise with \texttt{MESA} (version 9575; \citealt{Paxton:11, Paxton:13, Paxton:15}).  Our approach is similar to that of \citet{Paxton:11}, who also used \texttt{MESA} to model the evolution of the hydrostatic  layers during type I X-ray bursts.  We assume that the NS has a mass $M=1.4M_\odot$ and radius $R=10\trm{ km}$ and ignore corrections due to general relativity.

We assume pure He accretion (as in ultracompact X-ray binaries; \citealt{intZand:07}) since bursts that ignite in a pure He layer have especially high luminosities and strong PREs.  Systems that accrete H/He at mass accretion rates below $\approx 1\%$ of Eddington also ignite in a pure He layer and exhibit strong PREs \citep{Bildsten:98, Cumming:00, Galloway:17}. We assume that the atmosphere is always in local thermal equilibrium (LTE) and we model convection using mixing-length theory (MLT).  During the hydrostatic phase,  we set the top boundary at an optical depth of $\tau=100$ (during the wind phase we set the top boundary at a much smaller $\tau$ in order to capture the regions near the photosphere).  By neglecting the shallower layers, we avoid numerical difficulties while still being able to accurately follow the nuclear burning and the atmosphere's evolution.   In the Appendix \ref{app:mesa}, we provide our \texttt{MESA} inlist, which describes the setup we use in more detail.  

During the hydrostatic phase, it is convenient to parameterize the vertical coordinate in terms of the column depth $y$, defined as $y(r)=\int_r^{\infty}\rho dr$, where $\rho$ is the density and $r$ is the radius. Since the atmosphere is geometrically thin and in hydrostatic equilibrium up until the wind launches, $y\simeq P/g \simeq M_r/4\pi R^2$, where $P$, $g$, and $M_r$ are the pressure, gravitational acceleration, and mass above $r$, respectively. We simulate bursts for column depths at the ignition base ranging from $y_{\rm b}=3\times10^{8}-5\times10^{9}\,{\rm g\, cm^{-2}}$. The value of $y_b$ is controlled in \texttt{MESA} by varying the core luminosity and the accretion rate (the numerical settings are provided in the Appendix~\ref{app:mesa}; see \citealt{Cumming:03} and \citealt{Paxton:11} for a detailed description of the ignition model). 
We will primarily show results for three representative values: $y_{\rm b}=\left(0.5, 1.5, 5\right)\times10^{9} \ {\rm g\,cm^{-2}}$, which we will denote as (y1, y2, y3), respectively.  

We consider two reaction networks\footnote{\url{http://cococubed.asu.edu/code\_pages/burn.shtml}}: a simple 9-isotope network (\texttt{basic\_plus\_fe56.net}) denoted by n9, and a more complete 21-isotope network (\texttt{approx21.net}, which is based on the 19-isotope network by~\citealt{Weaver:78} with the extra inclusion of $^{56}$Fe and $^{56}$Cr) denoted by n21.  We primarily show results for y1n21, y2n21, and y3n21, but will sometimes also show the n9 variants in order to illustrate how the size of the reaction network can impact the simulations. 

In Sections \ref{sec:thermal_hse} and \ref{sec:composition_hse} we describe, respectively,  the thermal and compositional evolutions of the atmosphere during the rise.  Since the results are similar to those of \citetalias{Weinberg:06}, we only describe the key features of the rise and refer the reader to that paper for additional details.  It is worth noting, however, that they only consider relatively shallow ignition depths  of $y_b=(3-5)\times10^8\trm{ g cm}^{-2}$ compared to our $(3-50)\times10^8\trm{ g cm}^{-2}$.
 
\subsection{Evolution of the Thermal Profile}
\label{sec:thermal_hse}

In Figure~\ref{fig:convZone} we show the evolution of the thermal profile of model y3n21 during the burst rise.\footnote{In the X-ray burst literature, $y$ is usually plotted as increasing to the right.  We plot it as increasing to the left in order to match the orientation of the wind structure figures shown later, which are often plotted in terms of the radial coordinate $r$, rather than $y$.} As the base temperature $T_b$ rises due to He burning, a convective zone forms and begins to extend outward to lower pressure (smaller $y$) on a timescale of $\sim1 \trm{ ms}$ (for y1n21 it is about 50 times longer).   Initially, $T_b$ rises so quickly that there is not enough time for the radiative layer above the convective zone to thermally adjust.  As a result, the thermal profile in the radiative region is unchanged from the pre-ignition profile (see also \citetalias{Weinberg:06}, \citealt{Paxton:11}). This can be seen in Figure~\ref{fig:convZone} at times $t<-28.9\trm{ ms}$, where $t=0$ corresponds to when the wind turns on.  

Over most of the convective zone, the convection is highly subsonic and efficient  and the temperature profile very nearly follows an adiabat $T\propto y^n$, with $n\simeq 2/5$ (i.e., close to the adiabatic index of an ideal gas).  In the overlying radiative region, the temperature profile is shallower and since the opacity varies only slightly with column depth, $T\propto y^{1/4}$.

 \begin{figure}
\includegraphics[trim=-1.5cm 0 0 0.5cm, width=0.47\textwidth]{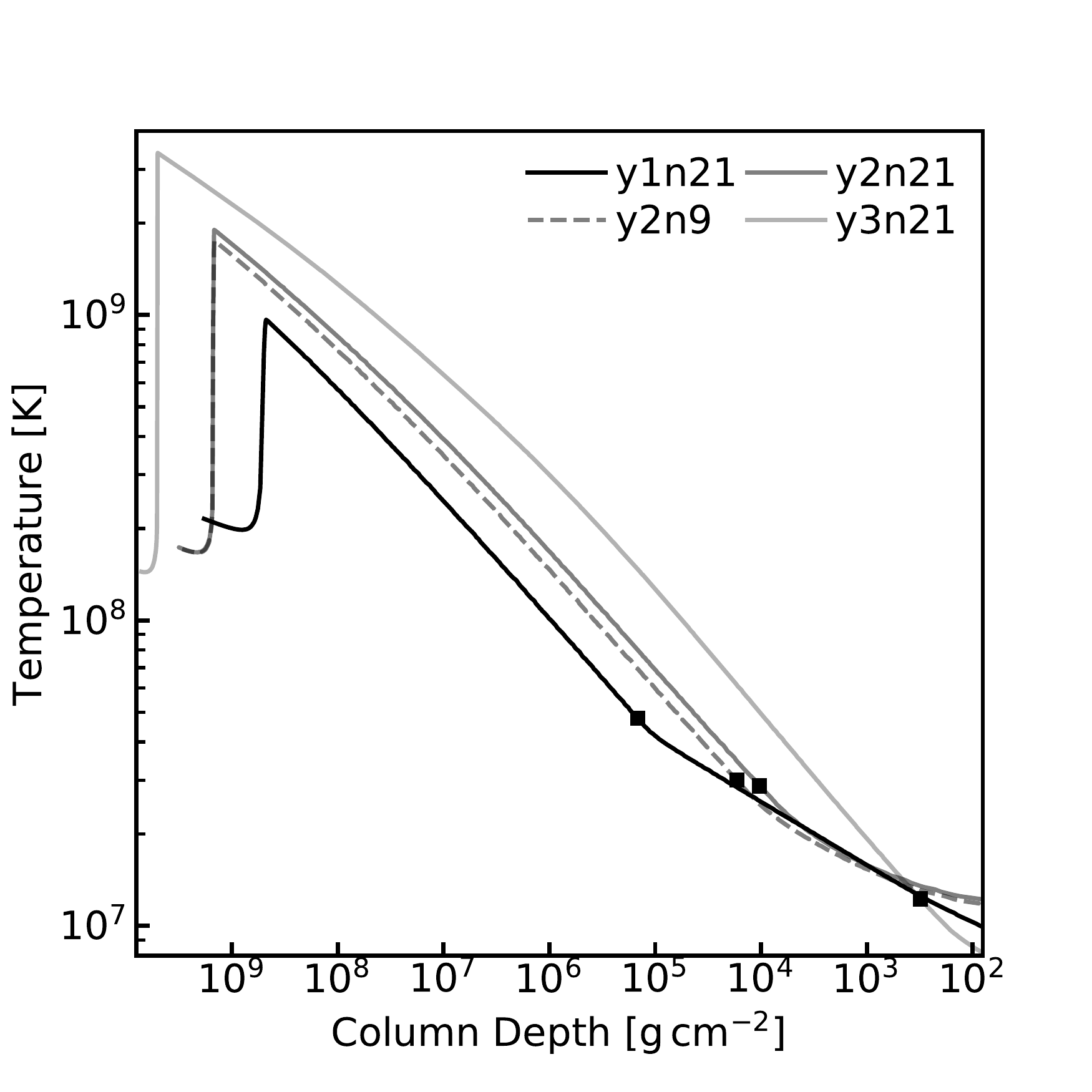}
\caption{Temperature as a function of column depth for five  of the burst models at the moment when their convective zones reach maximum extent.  The squares indicate the top of the convective zone. }
\label{fig:T_vs_y}
\end{figure}

 \begin{figure*}
\includegraphics[trim=4.1cm 0 0 0.5cm,width=1.10\textwidth]{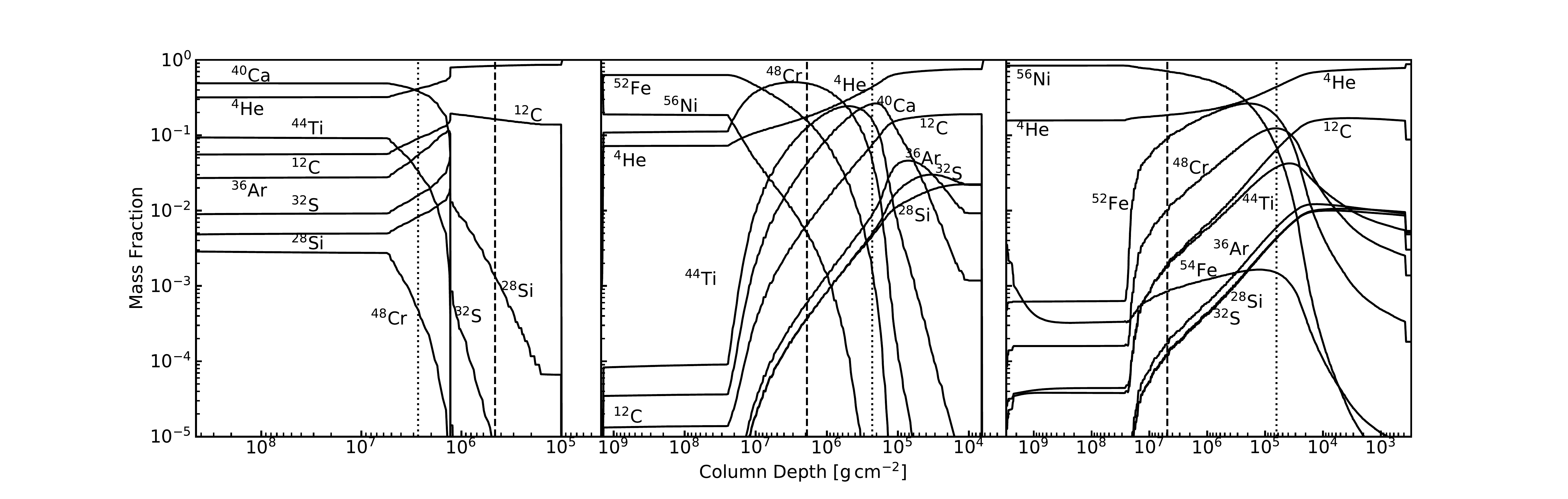}
\caption{Composition as a function of column depth at the moment just before the wind launches for models y1n21 (left), y2n21 (middle), and y3n21 (right). The dashed vertical lines indicate $y_{\rm wb}$, the maximum column depth of the wind base.  The dotted vertical lines indicate  $y_{\rm ash}$, the location where the mass fraction of heavy elements ($A>40$) equals 50\%.}
\label{fig:comp_init}
\end{figure*}

Eventually, the top of the convective zone reaches low enough $y$ that the local thermal timescale of the overlying radiative layer equals the heating timescale at the base. The radiative flux can then diffuse outward through the radiative region without being overtaken by the growing convective region.  This flux begins to heat the radiative region and the convective zone gradually retreats downward to larger $y$ (Figure~\ref{fig:convZone} at $t>-28.9\trm{ ms}$).  As the radiative region heats up, the radiative luminosity $\Lrad$ in the shallower layers begins to approach the local Eddington limit
\beq
L_{\rm Edd}(T) = \frac{4\pi c GM}{\kappa(T)}.
\label{eq:Ledd}
\eeq
The opacity $\kappa(T)$ is dominated by electron scattering and is temperature-dependent due to Klein-Nishina (i.e., special relativistic) corrections.  It varies approximately as (\citealt{Paczynski:83}; \texttt{MESA} uses a more exact form)
\beq
\kappa(T)=\kappa_0\left[1+\left(\frac{T}{0.45\trm{ GK}}\right)^{0.86}\right]^{-1},
\label{eq:kappa_vs_T}
\eeq
where $\kappa_0\simeq 0.2 (1+X_{\rm H}) \trm{ cm$^2$ g$^{-1}$}$ and $X_{\rm H}$ is the hydrogen mass fraction.  
Since $\kappa$ is larger for smaller $T$,  smaller $y$ have smaller $L_{\rm Edd}$.  As a result, a given luminosity $L$ can be sub-Eddington in the deep hotter layers, but becomes super-Eddington as the radiation diffuses upward into the shallow cooler layers \citep{Ebisuzaki:83}.  Indeed, as we will show in the hydrodynamic simulations (see Section \ref{sec:wind_profiles} and Figure \ref{fig:L_vs_y_y2n21}), the base of the wind is initially at small $y$, moves to larger $y$ as the deeper layers heat up, and finally moves back outward to smaller $y$ as the layers begin to cool.  

We run the hydrostatic simulations until the moment the luminosity first exceeds the local Eddington limit (defined as $t=0$).  As can be seen in Figure~\ref{fig:convZone}, at $t=0$ the convective zone has retreated and the atmosphere is almost fully radiative.

\subsection{Pre-wind Composition Profile}
\label{sec:composition_hse}

As the convective zone extends outward, it efficiently mixes the ashes of burning up to lower column depths.  The minimum column depth $y_{c, \rm min}$ reached by the convective zone, and hence reached by the ashes of burning, is shown by the solid squares in Figure~\ref{fig:T_vs_y} for five of the burst models.  For model y3n21, which has the deepest ignition and thus the largest energy release, $y_{c,\rm min}\approx 10^3\trm{ g cm}^{-2}$, while for the other models, $y_{c, \rm min}\approx 10^4-10^5\trm{ g cm}^{-2}$ (consistent with  \citetalias{Weinberg:06}).  We also see that a more complete reaction network (n21 compared to n9) results in slightly smaller $y_{c, \rm min}$ due to the increased energy release; comparing the y2 and y3 models, we find that this difference becomes more significant at larger ignition depths.

In Section \ref{sec:num_wind} we describe the time-dependent wind and show that  the column depth of the wind's base $y_{\rm wb} \gg y_{c, \rm min}$. As a result, ashes are ejected by the wind and exposed.  In Figure~\ref{fig:comp_init} we show the composition profiles of models y1n21, y2n21, and y3n21 at the end of the hydrostatic phase, just before the wind is launched. 
The dashed vertical lines indicate $y_{\rm wb}$.  At a given $y$, the composition is determined by the burning stage at the moment $y_c(t)=y$, where $y_c(t)$ is the location of the top of the retreating convective zone.   
For y1n21 ($y_{\rm b}=5\times10^8\,{\rm g\,cm^{-2}}$), we see that the wind will be dominated by light elements, primarily $^{4}$He with a small amount of $^{12}$C.
However, for y2n21 ($y_{\rm b}=1.5\times10^9\,{\rm g\,cm^{-2}}$), the wind will be dominated by heavy elements such as $^{48}$Cr and $^{52}$Fe, while for models ignited at even deeper depths (y3n21; $y_{\rm b}=5\times10^9\,{\rm g\,cm^{-2}}$)  the wind will be primarily $^{56}$Ni.\footnote{\cite{Woosley:04} found that bursts can dredge up the ashes of previous bursts. We do not include such ashes in our simulations and instead focus on newly synthesized elements. We therefore assume that for $y>y_b$, the composition is pure $^{56}$Cr, the end product of the 21-isotope network.  This ensures that the mass fractions of elements like $^{56}$Ni and $^{54}$Fe in the wind are  not the result of having been dredged up by convection.  We find that  $^{56}$Cr has a mass fraction of $10^{-4}-10^{-3}$ for $y_b<y<y_{c, {\rm min}}$ due to dredge-up (not shown in Figure~\ref{fig:comp_init}).\label{fn:dredgeup}} 

\section{Hydrodynamic wind}
\label{sec:num_wind}

When the luminosity first exceeds the local Eddington limit $L_{\rm Edd}$ (Equation \ref{eq:Ledd}), we stop the hydrostatic calculation. We use the last hydrostatic profile as the initial conditions for the time-dependent spherically symmetric hydrodynamic equations, which we integrate using \texttt{MESA}'s implicit hydrodynamics solver. The \texttt{MESA} inlist for our hydrodynamic calculations is given in the Appendix \ref{app:mesa}.  Since the atmosphere is almost fully radiative at this stage, we turn off MLT (see \citealt{Ro:16} and \citealt{Quataert:16} for a discussion of  convective stability in radiation-driven winds).  
We include radiation in the diffusion approximation ($dT^4/dr=-3\kappa \rho \Lrad/4\pi ac r^2$)  and set the upper boundary at  optical depth  $\tau=1$. We define the photospheric radius $r_{\rm ph}$ as the location where $\Lrad/4\pi r_{\rm ph}^2=\sigma T^4$ (similar to \citealt{Quinn:85} and \citealt{Paczynski:86}).  In practice, we find that the optical depth at $r_{\rm ph}$ is $\tau = \int_{r_{\rm ph}}^\infty \kappa \rho dr \approx 3$. Thus, for $r < r_{\rm ph}$ the diffusion approximation and LTE should be valid.   In the region between $r_{\rm ph}$ and the upper boundary of our grid ($\tau=1$) deviations from LTE may occur, although we expect the photons and gas particles to still be well coupled \citep{Joss:87}.  Nonetheless, our results should be treated as approximations of the true structure in this region (see the steady-state models of \citealt{Joss:87} and \citealt{Nobili:94} for more detailed treatments of this region and the optically thin region above it). Finally, to account for the mass-loss at the top of our grid, we repeatedly remove the top layer of the atmosphere when its density drops below a threshold value of $10^{-14}\,{\rm g\,cm^{-3}}$ (by experimenting with different thresholds, we determined that the wind solution is not affected by this procedure).

We describe the evolution of the wind structure in Section~\ref{sec:wind_profiles}.  We compare our results with steady-state models in Section~\ref{sec:steady_state} and use these results in Section~\ref{sec:wind_vs_yb} to explain why the wind structure is not sensitive to ignition depth.  Finally, in Section~\ref{sec:ejection_of_heavy_elements} we describe the composition of the wind.

\subsection{Time-dependent Wind Profiles}
\label{sec:wind_profiles}

\begin{figure}
\includegraphics[trim=-0.9cm 0 0 0.5cm,width=0.45\textwidth]{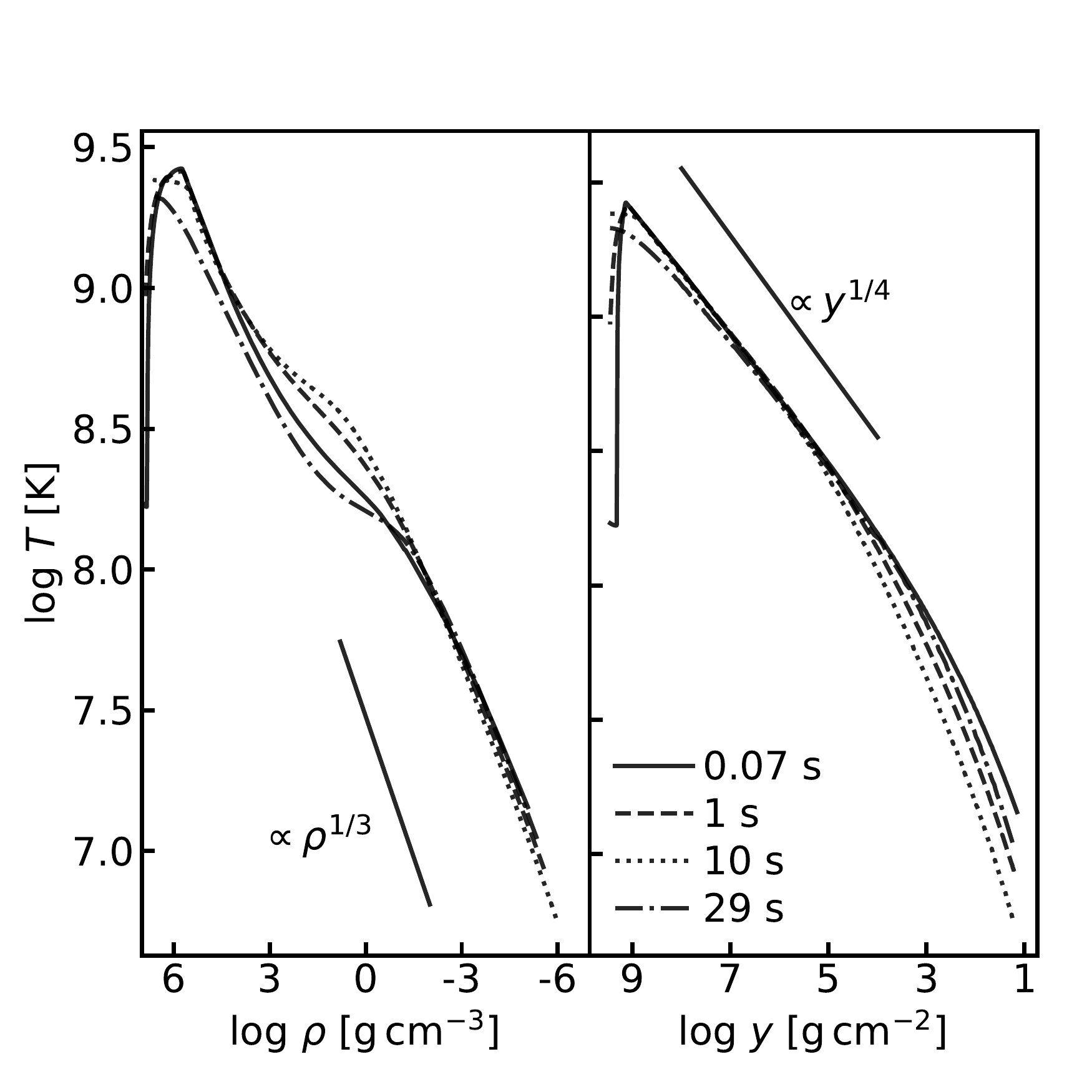}
\caption{Temperature as a function of density (left panel) and column depth (right panel) for model y2n21 at different times during the wind phase.}
\label{fig:T_vs_rho_y2n21}
\end{figure}

\begin{figure}
\includegraphics[trim=-0.9cm 0 0 0.5cm,width=0.45\textwidth]{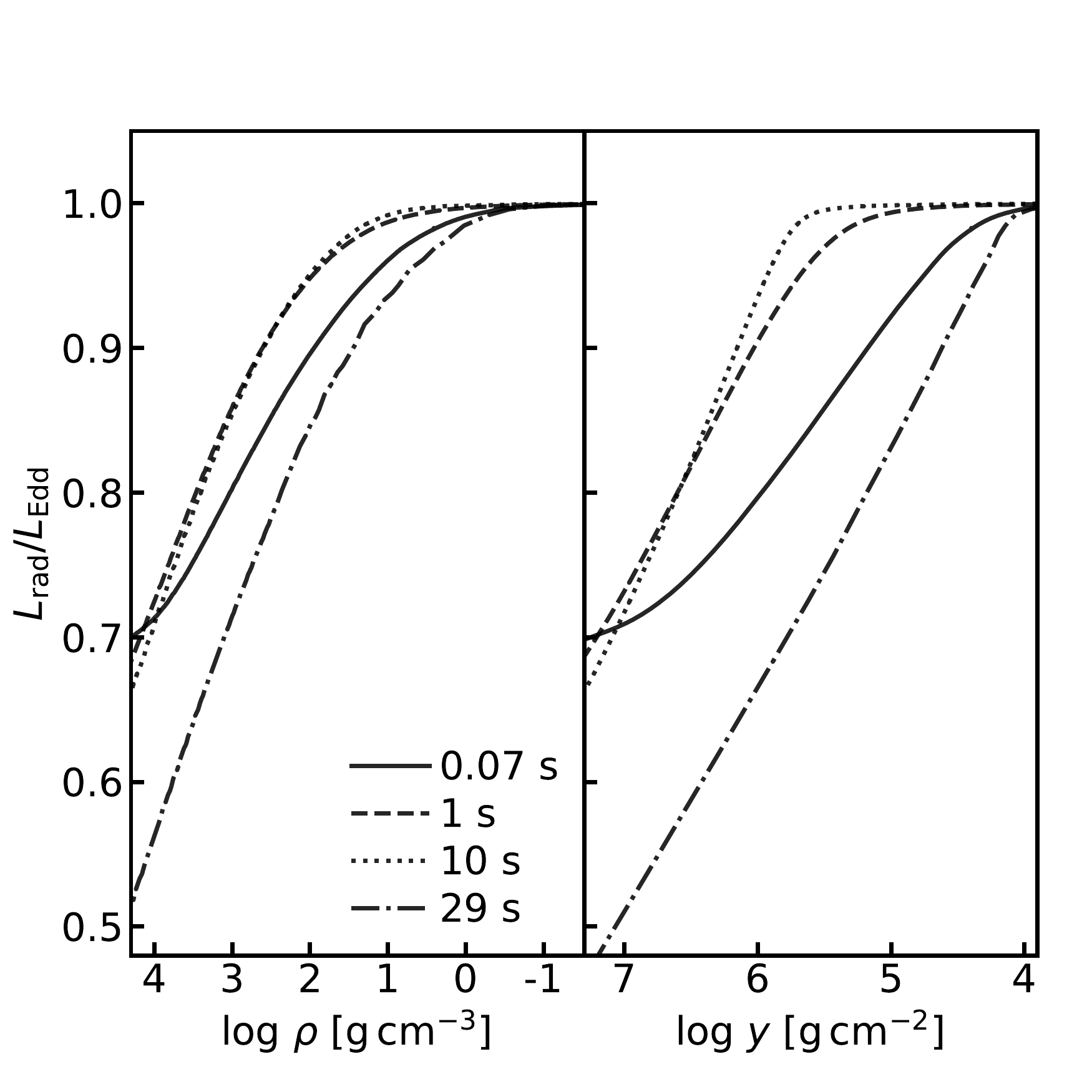}
\caption{Luminosity $\Lrad$ relative to Eddington $\Ledd$ as a  function of density (left panel) and column depth (right panel) for model y2n21 at different times during the wind phase. }
\label{fig:L_vs_y_y2n21}
\end{figure}

In Figures~\ref{fig:T_vs_rho_y2n21} and \ref{fig:L_vs_y_y2n21} we show  the temperature and luminosity  as a function of both density and column depth at four different times during the hydrodynamic wind phase of model y2n21.    The location where $\Lrad$ first exceeds $L_{\rm Edd}$ corresponds to the wind base. Note that $L$ never exceeds $L_{\rm Edd}$ by more than a few percent, since any excess luminosity is used to expel matter to infinity \citep{Ebisuzaki:83,Kato:83,Paczynski:86}.  At early times ($t=0.07\trm{ s}$), the column depth of the wind base $y_{\rm wb}\approx 10^4\trm{ g cm}^{-2}$.  As the wind evolves during the next $\approx 10\trm{ s}$,  the location where $\Lrad>L_{\rm Edd}$ moves to larger $y$ and $\rho$ and thus higher $T$, eventually reaching as far down as $y_{\rm wb}\simeq 10^6\trm{ g cm}^{-2}$.    By $t=29\trm{ s}$, the NS surface layers have cooled, $y_{\rm wb}$ has moved back to shallower depths, and the wind dies down.   As we explain in Section~\ref{sec:steady_state}, the profiles at depths greater than that of the wind base approximately follow power-law relations $T\propto\rho^{1/3}\propto y^{1/4}$. The $T\propto \rho^{1/3}$ relation also holds in regions sufficiently above the wind base.

\begin{figure}
\includegraphics[trim=-0.9cm 0 0 -0.5cm,width=0.47\textwidth]{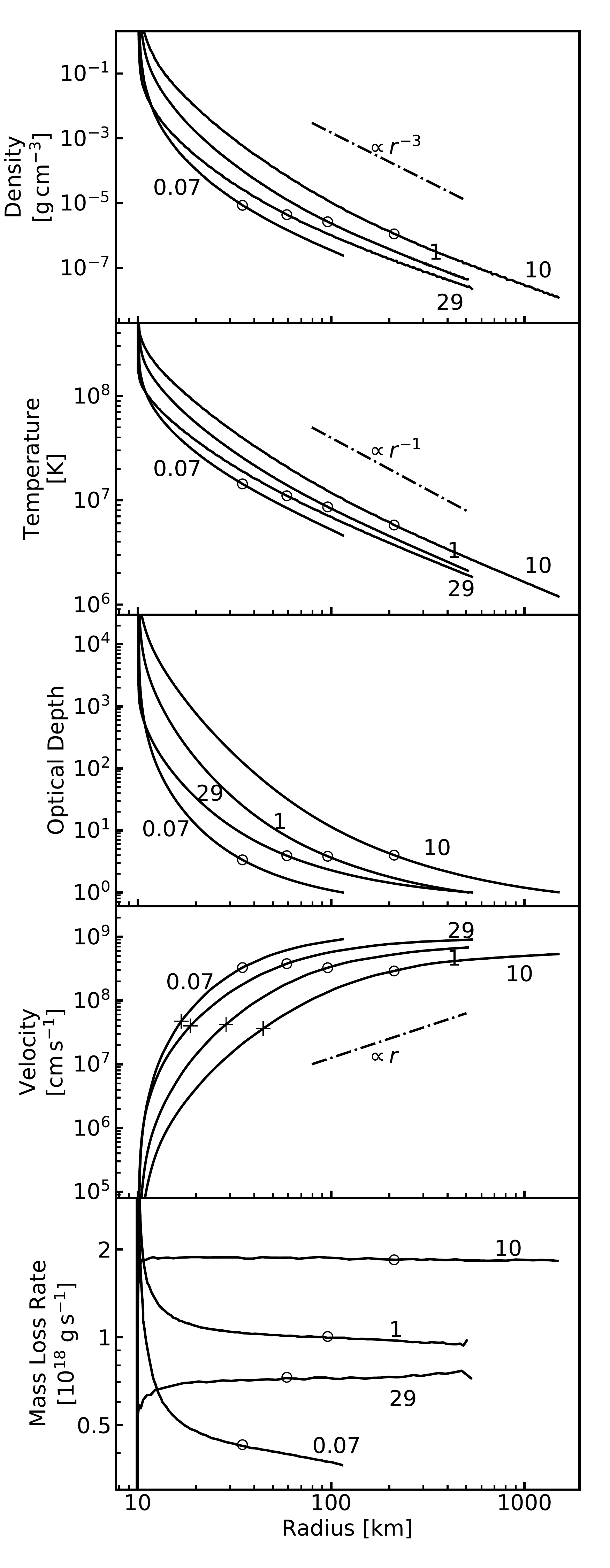}
\caption{Radial profiles of the wind structure at different times for model y2n21. The numbers mark the time (in seconds) since wind onset. Each curve is terminated at the location where the optical depth $\tau=1$. The circles indicate the location of the photosphere $r_{\rm ph}$. In the velocity-radius plot (fourth panel), the pluses  indicate the location of the isothermal sonic point.  
}
\label{fig:wind_pf_y2n21}
\end{figure}

In Figure~\ref{fig:wind_pf_y2n21} we show the wind structure of model y2n21 in more detail.  We plot profiles as a function of $r$ rather than $y$ or $\rho$ in order to more clearly reveal the structure of the tenuous outer regions of the wind out to $r\sim 10^3\trm{ km}$. The open circles indicate the location of the photosphere $r_{\rm ph}$.  We see that there is a large radius expansion, with the photosphere reaching a maximum of $r_{\rm ph}\simeq 200\trm{ km}$ (see also Figure~\ref{fig:wind_hist}). The pluses indicate the location of the isothermal sonic point, defined as the radius where the velocity satiesfies $v^2=kT/\mu m_p$, where $\mu$ is the mean molecular weight and $m_p$ is the proton mass (see, e.g., \citealt{Quinn:85, Joss:87}); the equilibrium sonic point where $v^2=dP/d\rho$ occurs at $\tau < 1$ and is thus beyond our simulated region. 

As the wind gains strength during the first 10 seconds, the mass-loss rate $\Mdot$, temperature, density, and optical depth all increase throughout the wind.  The velocity, which never exceeds $\sim 0.01c$,  decreases during this time, since $\Mdot\simeq 4\pi r^2\rho v$ only changes by order unity whereas $\rho$ increases significantly. At $t\simeq10\trm{ s}$ the wind settles into a steady state and for the next $\approx 15\trm{ s}$ the profiles change very little. For model y2n21, $\Mdot\simeq 2\times10^{18}\trm{ g s}^{-1}$ at its maximum. By $t=29\trm{ s}$, the energy and mass supply have dwindled and $\Mdot$  decreases.  As a result, the temperatures and densities drop and the photosphere begins to fall back to the NS  surface.  

\begin{figure}
\includegraphics[trim=0.0cm 0 0 0.5cm, width=0.46\textwidth]{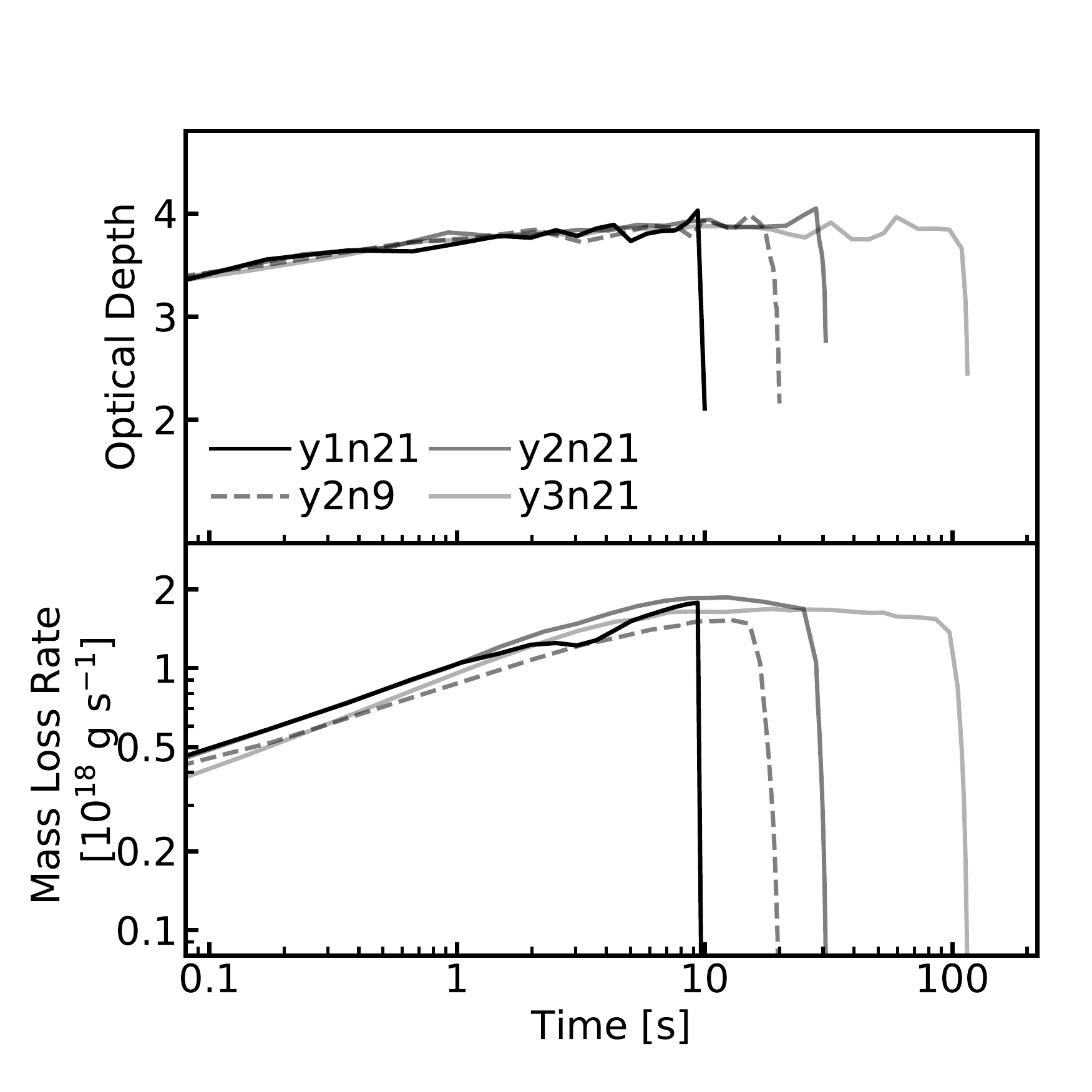}
\caption{Optical depth (top panel) and mass-loss rate (bottom panel) at the photosphere as a function of time for four of the burst models.}
\label{fig:wind_hist_ph_supp}
\end{figure}

We find that aside from differences in duration, the wind profiles of our other burst models are all similar to that of model y2n21 despite the  significant range of ignition depth.  This is because the wind structure is largely determined by $\Mdot$  \citep{Kato:83, Quinn:85, Paczynski:86}, and the different models all have very similar $\Mdot(t)$ up until the wind terminates.  We illustrate this in the bottom panel of Figure~\ref{fig:wind_hist_ph_supp} for four of the models.  In Section~\ref{sec:wind_vs_yb}, we explain why $\Mdot$ is a weak function of $y_b$.

In the top panel of Figure~\ref{fig:eta_vs_yb} we show the ratio of the total mass ejected by the wind $M_{\rm ej}$ at the end of the PRE to the accreted mass $M_{\rm accr}\simeq 4\pi R^2 y_b$.   We find
\beq
\label{eq:eta}
\eta \equiv \frac{M_{\rm ej}}{M_{\rm accr}} \simeq 2.5\times10^{-3}
\eeq
almost independent of $y_b$.  Burst energetics set an upper bound of $\eta \la 8\times 10^{-3}$, which is given by the ratio of the nuclear energy release per nucleon $\simeq 1.6\trm{ MeV nucleon}^{-1}$ (i.e., the difference in binding energy  between $^4$He and $^{56}$Ni) to the gravitational binding energy per nucleon $GM/R\simeq 200\trm{ MeV nucleon}^{-1}$.  The value $\eta\simeq 2.5\times10^{-3}$ implies that $\approx 30\%$ of the nuclear energy goes to unbinding matter from the NS, independent of ignition depth.

\subsection{Comparison with Steady-state Models}
\label{sec:steady_state}

Since the flow is subsonic at radii smaller than the equilibrium sonic point (which is located  at an optical depth $\tau<1$), the structure throughout the modeled region is nearly in hydrostatic equilibrium at each instant.  Therefore, the evolution approximately follows a sequence of steady-state solutions (i.e., quasi-static profiles) determined by the instantaneous $\Mdot(t)$. Indeed, our profiles are qualitatively similar to those of  steady-state wind models in which $\Mdot$ is treated as a free parameter \citep{Ebisuzaki:83, Kato:83, Paczynski:86, Joss:87}.  In steady-state, the time-dependent terms vanish and the mass, momentum, and energy equations are
\begin{eqnarray}
&& \Mdot =4\pi r^2 \rho v = {\rm constant}, \label{eq:mass_cons}\\
&&v\frac{\diff v}{\diff r} = -\frac{1}{\rho}\frac{\diff P}{\diff r} - g, \label{eq:momentum_cons}\\
&&\Edot=\Mdot \left(\frac{v^2}{2} - \frac{GM}{r} + h \right) + \Lrad =  {\rm constant}, \label{eq:E_cons}
\end{eqnarray}
where $\Edot$ is the energy-loss rate of the wind, $h=(U+P)/\rho$ is the enthalpy, and $U$ is the energy density.  Over a large region between the wind base and the equilibrium sonic point, we find that $dP/dr\simeq -\rho g$, radiation pressure dominates so that $P \propto T^4$ and $h\simeq 4P/\rho$, and $\Lrad(r\gg R)\simeq L_{\rm Edd, 0}\simeq \Edot$, where $L_{\rm Edd,0}=4\pi cGM/\kappa_0$.  Together these imply that over this region $P\propto \rho^{4/3}$ and the fluid behaves as if it has an adiabatic index $\gamma=4/3$, as also noted by \citet{Kato:86}.\footnote{Although the photon diffusion time $t_{\rm diff}\simeq r^2\kappa \rho /c$ is much shorter than the advective time $t_{\rm adv}\simeq r/v$, and thus heat flows in and out of a fluid element, the entropy profile of the wind is nearly constant over a large region \citep{Kato:86}.}  As a result, $\rho \propto r^{-3}$, $T \propto r^{-1}$, and $v \propto r$, as shown in  Figures~\ref{fig:T_vs_rho_y2n21} and \ref{fig:wind_pf_y2n21} (see also Figure~5 in \citealt{Paczynski:86}). 

Given that $\rho\propto r^{-3}$, the optical depth $\tau \simeq \tau^\ast/2$, where $\tau^\ast=\kappa\rho r$ is the effective optical depth used in the steady-state wind calculations of \citet{Quinn:85} and \citet{Paczynski:86}.  In Figure~\ref{fig:wind_hist_ph_supp} we show that at the photosphere $\tau\simeq 3.5$ nearly independent of time and ignition model, which is comparable to the values found by  \citet{Quinn:85} and \citet{Paczynski:86}.

Since $\Lrad$ is much larger than the kinetic power,  X-ray burst winds are in the opposite regime from massive star winds, which \citet{Quataert:16} studied.  Their analytic steady-state model is therefore not directly applicable here. More recently, \citet{Owocki:17} derived semi-analytic steady-state wind solutions that bridge the two regimes.   Although we have not attempted to implement their solutions, they should be applicable to the steady-state regime of PRE  bursts.

\subsection{Dependence of Wind Structure on Ignition Depth}
\label{sec:wind_vs_yb}

\begin{figure}
\includegraphics[trim=-0.9cm 0 0 0.0cm,width=0.47\textwidth]{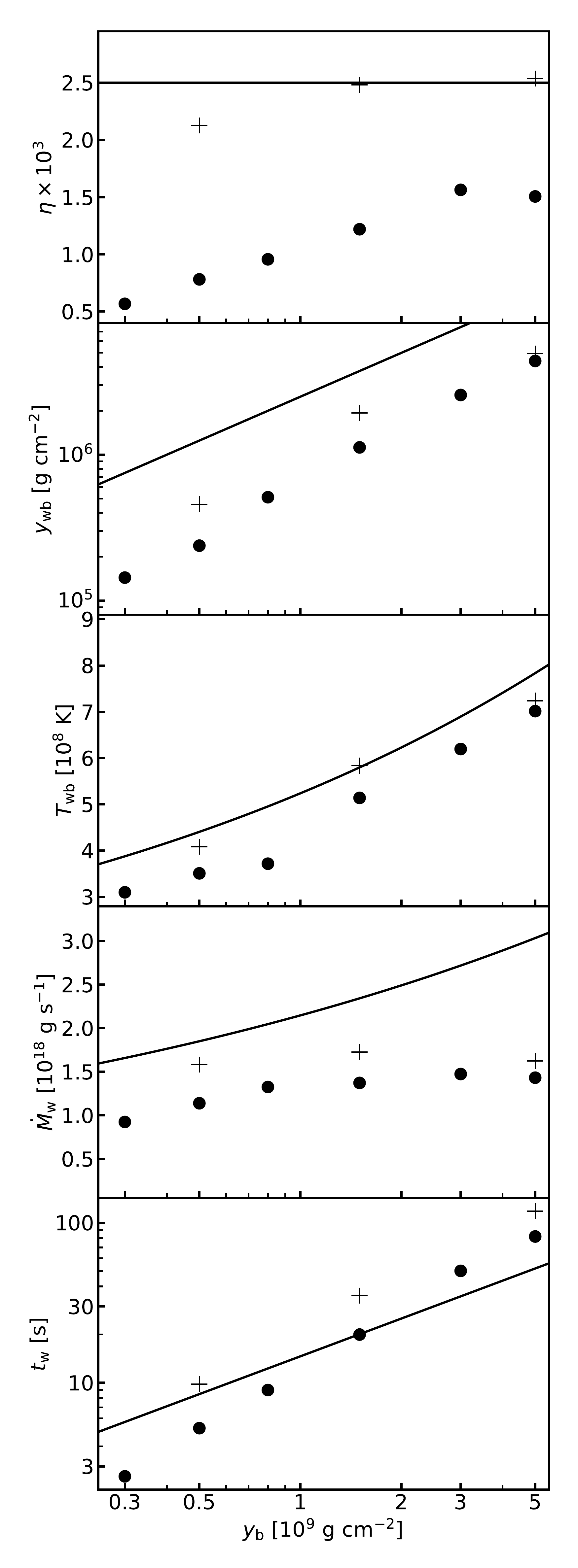}
\caption{
Bulk properties of the wind as a function of ignition column depth $y_{\rm b}$. The plus (circle) symbols are from numerical simulations using the 21-isotope (9-isotope) network. The $y_{\rm wb}$, $T_{\rm wb}$, and $\Mdot$ panels give the values during the approximate steady-state wind phase. The curves are the analytic approximations described in the text.}
\label{fig:eta_vs_yb}
\end{figure}

The duration of the wind increases with ignition depth $y_b$, but its structure is nearly independent of $y_b$.  This is because  $\dot{M}_{\rm w}(t)$, which effectively sets the wind structure (see, e.g., \citealt{Kato:83, Paczynski:86}), depends only weakly on $y_b$. We can understand this weak dependence by appealing to energy conservation, assuming a steady-state wind.  Just above the base of the wind ($r_{\rm wb}\simeq R$), the enthalpy and kinetic energy are small compared to the binding energy,  and by Equation~(\ref{eq:E_cons}) 
\begin{equation}
\Edot \simeq -\frac{GM \Mdot}{R} + \Lrad(r_{\rm wb}),
\label{eq:Edot_wb}
\end{equation} 
where $\Lrad(r_{\rm wb})\simeq 4\pi c G M/\kappa_{\rm wb}$, i.e., the  Eddington luminosity at the wind base, with $\kappa_{\rm wb}=\kappa\left[T(r_{\rm wb})\right]$.  At $r\gg R$, the flow of mechanical energy is small compared to $L_{\rm rad}$, and 
\beq
\Edot\simeq L_{\rm rad}(r\gg R) \simeq \frac{4\pi c GM}{\kappa_0},
\label{eq:Edot_infty}
\eeq
where the last equality follows because the luminosity at large $r$ only slightly exceeds the local Eddington limit. During the steady state, $\Edot$ is constant throughout the wind, and we can equate Equations (\ref{eq:Edot_wb}) and (\ref{eq:Edot_infty}) to find
\beq
\Mdot \simeq \frac{4\pi c R}{\kappa_0}\left[\frac{\kappa_0}{\kappa_{\rm wb}}-1\right]\simeq  \frac{4\pi c R}{\kappa_0}\left(\frac{T_{\rm wb}}{0.45\trm{ GK}}\right)^{0.86}
\label{eq:Mdot_Twb}
\eeq
(\citealt{Paczynski:86} derive a similar expression). We now show that $T_{\rm wb}$ (and hence $\Mdot$) is a weak function of $y_b$ by first estimating the peak temperature at the ignition base $T_b(y_b)$ and then relating $T_b$ to $T_{\rm wb}$.

\begin{figure}
\includegraphics[trim=-1.0cm 0 0 0.5cm, width=0.46\textwidth]{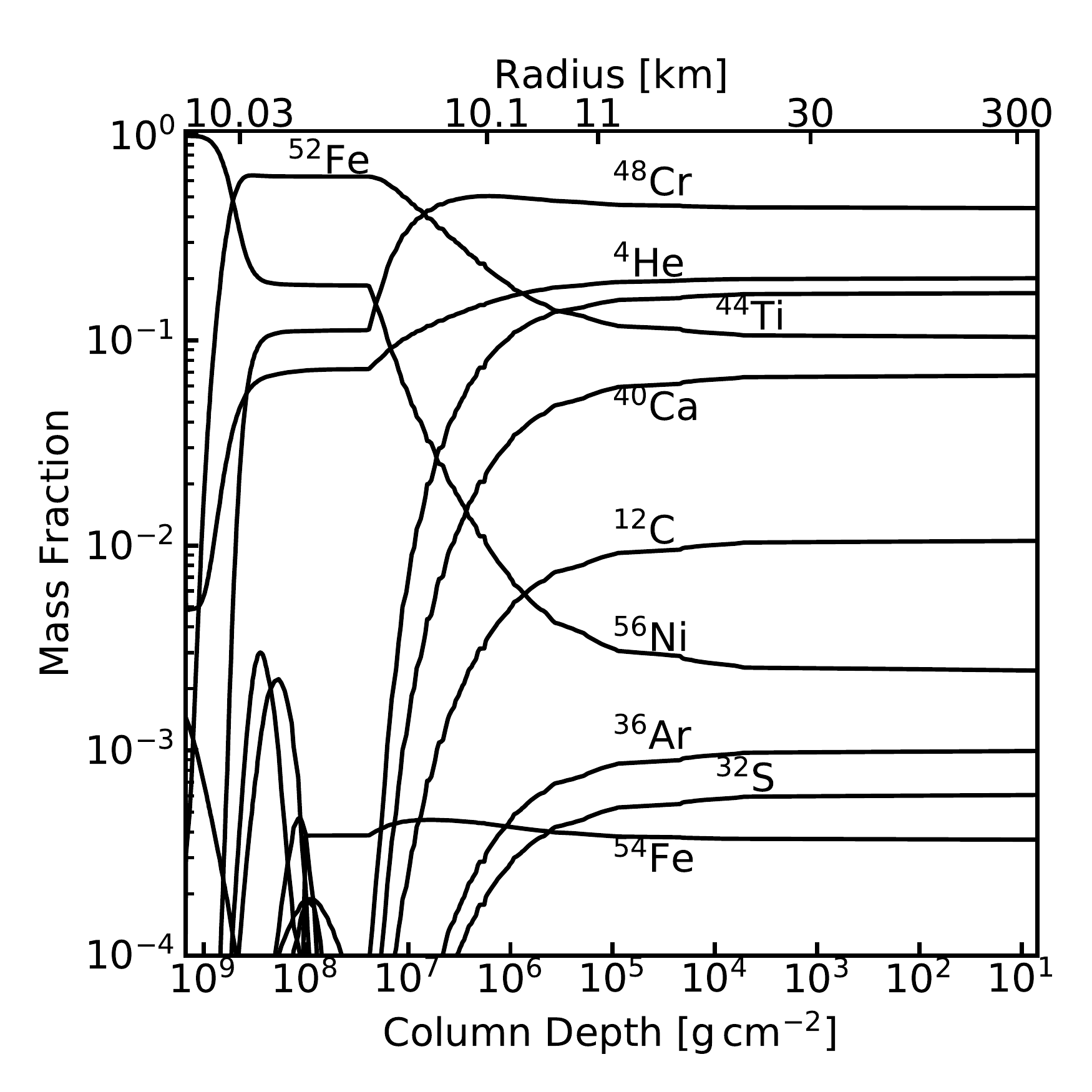}
\caption{Composition of the wind as a function of column depth (bottom axis) and radius (top axis) at $t=10\,{\rm s}$ for model y2n21. The wind is in a near steady state by this time.}
\label{fig:comp_y2n21}
\end{figure}

The base temperature $T_b$ rises until it becomes radiation-pressure-dominated, which lifts the degeneracy and stifles the burning \citep{Fujimoto:81, Bildsten:98}. At its maximum,
\beq
T_b\simeq f\left(\frac{3 g y_b}{a}\right)^{1/4} \simeq 2.3 \, y_{b,9}^{1/4}\trm{ GK},
\eeq
where $y_{b,9}=y_b/10^9\trm{ g cm}^{-2}$ and at $f=1$ radiation pressure completely dominates.  In the numerical expression here and below we set $f=0.8$ based on our numerical calculations (see Figure~\ref{fig:T_vs_rho_y2n21} and also \citetalias{intZand:10}).  During the wind phase, the bound layers are fully radiative and satisfy $T\propto y^{1/4}$. When the wind is at its peak strength, $y_{\rm wb} \simeq \eta y_b$ and $T_{\rm wb} \simeq \eta^{1/4}T_b$, where $\eta$ is given by Equation (\ref{eq:eta}).  We thus find
\bea
\label{eq:ywb}
y_{\rm wb} &\simeq& 2.5\times10^6 \,y_{b,9} \trm{ g cm}^{-2},\\
\label{eq:Twb}
T_{\rm wb} &\simeq&  0.5\, y_{b,9}^{1/4}\trm{ GK},
\eea
where here and below we set $\eta=2.5\times10^{-3}$  (see top panel of Figure~\ref{fig:eta_vs_yb}).  In practice, this leads to a slight overestimate of $y_{\rm wb}$ since $\eta$ is determined by the total ejected mass at the end of the burst and therefore $\eta \ga y_{\rm wb}/y_b$.   Plugging Equation (\ref{eq:Twb}) into Equation (\ref{eq:Mdot_Twb}), we find that during the approximate steady-state phase
\beq
\Mdot \simeq 2.1\times10^{18} y_{b,9}^{0.22}\trm{ g s}^{-1}.
\label{eq:Mdot}
\eeq
As we show in Figure~\ref{fig:eta_vs_yb}, this estimate agrees reasonably well with the wind simulation results (the simulations show a somewhat smaller $\Mdot$ and an even weaker $y_b$ dependence).  We thus see that $\Mdot$ and therefore the wind structure is nearly independent of $y_b$.

Given $\Mdot$, we can estimate the wind duration
\beq
t_{\rm w} = \frac{M_{\rm ej}}{\dot{M}}=15\,y_{b,9}^{0.79}\trm{ s}.
\label{eq:tw}
\eeq
This compares well with the numerically calculated value of the total wind duration, shown in the bottom panel of Figure~\ref{fig:eta_vs_yb}. The latter is slightly larger because Equation (\ref{eq:Mdot}) overestimates $\Mdot$, especially near the beginning and end of the wind.

\subsection{Ejection of Heavy Elements}
\label{sec:ejection_of_heavy_elements}

In Figure~\ref{fig:comp_y2n21} we show the wind composition as a function of radius (top axis) and column depth (bottom axis) for model y2n21 at $t=10\trm{ s}$; by this time, the wind has settled into a steady state.  We find that the wind at that time is dominated by heavy-element ashes, particularly $^{48}$Cr, $^{44}$Ti, and $^{52}$Fe, whose mass fractions are about $0.5$, $0.2$  and $0.1$, respectively.  Comparing with the pre-wind profile, we see that this is the material that initially resided at a column depth $y\simeq 10^6\trm{ g cm}^{-2}$ (see dashed vertical line in middle panel of Figure~\ref{fig:comp_init}).  This is because $\Mdot\approx 10^{18}\trm{ g s}^{-1}$ and thus after $t=10\trm{ s}$, the wind has ablated the surface layers down to a depth $\Mdot t/4\pi R^2 \approx 10^6\trm{ g cm}^{-2}$. 

\begin{figure}
\includegraphics[trim=-0.4cm 0 0 0.5cm, width=0.44\textwidth]{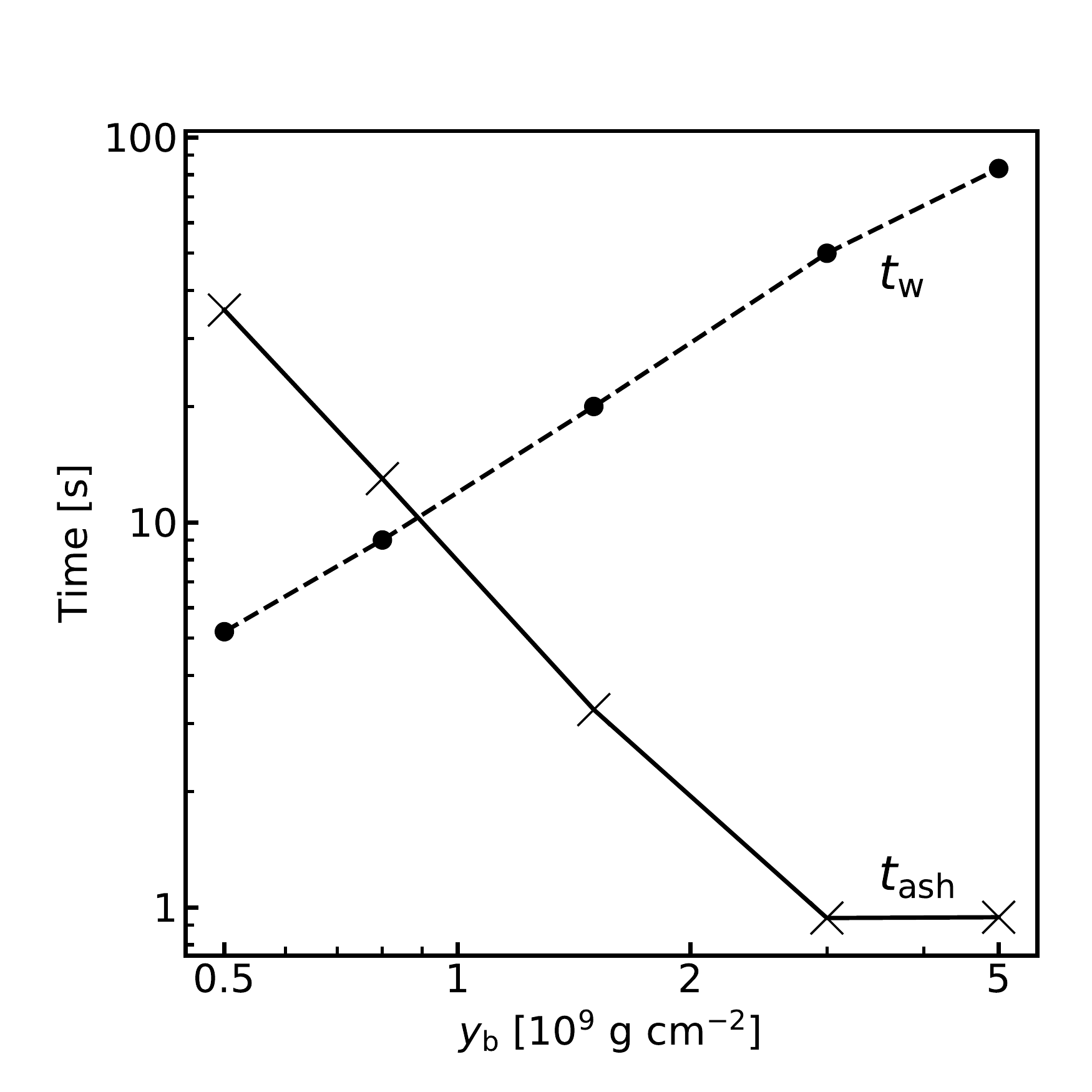}
\caption{Timescale to expose the heavy-element ashes $t_{\rm ash}$ (crosses) and the duration of the wind $t_{\rm w}$ (circles) as a function of ignition column depth $y_b$.  The $t_{\rm w}$ points are from the n9 models (see also the bottom panel of Figure~\ref{fig:eta_vs_yb}). For clarity, we connect the points with straight lines.}
\label{fig:t_ash_vs_y_b}
\end{figure}

An interesting feature of the pre-wind profile is that  for deep ignitions ($y_{b,9}=1-5$) the column depth at which the composition transitions from mostly light to mostly heavy elements is almost a constant value of $y_{\rm ash}\approx 10^5\trm{ g cm}^{-2}$. In Figure~\ref{fig:comp_init} we indicate $y_{\rm ash}$ with a vertical dotted line, where we formally define $y_{\rm ash}$ as the location  where the total mass fraction of elements with mass number $A>40$ equals 0.5 (for $y<y_{\rm ash}$, the elements consist predominantly of $^4$He and $^{12}$C). We can then define the timescale to expose the heavy ashes
\begin{eqnarray}
t_{\rm ash}&=& \frac{4\pi R^2 y_{\rm ash}}{\dot{M}_{\rm w}}\nonumber \\
&\simeq&\left(1.3\,{\rm s}\right) \left(\frac{y_{\rm ash}}{10^5\,{\rm g\, cm^{-2}}}\right)\left(\frac{10^{18}\,{\rm g\,s^{-1}}}{\dot{M}_{\rm w}}\right).
\end{eqnarray}
In Figure~\ref{fig:t_ash_vs_y_b} we plot $t_{\rm ash}$ as a function of $y_{\rm b}$. 
Starting from $y_{b,9}=0.5$, we find that  $t_{\rm ash}$ first decreases sharply as $y_b$ increases,  but then for  $y_{b,9}\gtrsim 3$ it plateaus at $t_{\rm ash}\simeq 1\trm{ s}$.  This is because $y_{\rm ash}$ plateaus at $y_{\rm ash}\approx 10^5\trm{ g cm}^{-2}$, while $\Mdot$ depends very weakly on $y_b$ (see Section \ref{sec:wind_vs_yb}).  In Section~\ref{sec:comp_pre}, we describe how this might explain a feature of superexpansion bursts.

\begin{figure}
\includegraphics[trim=-0.4cm 0 0 0.5cm,width=0.45\textwidth]{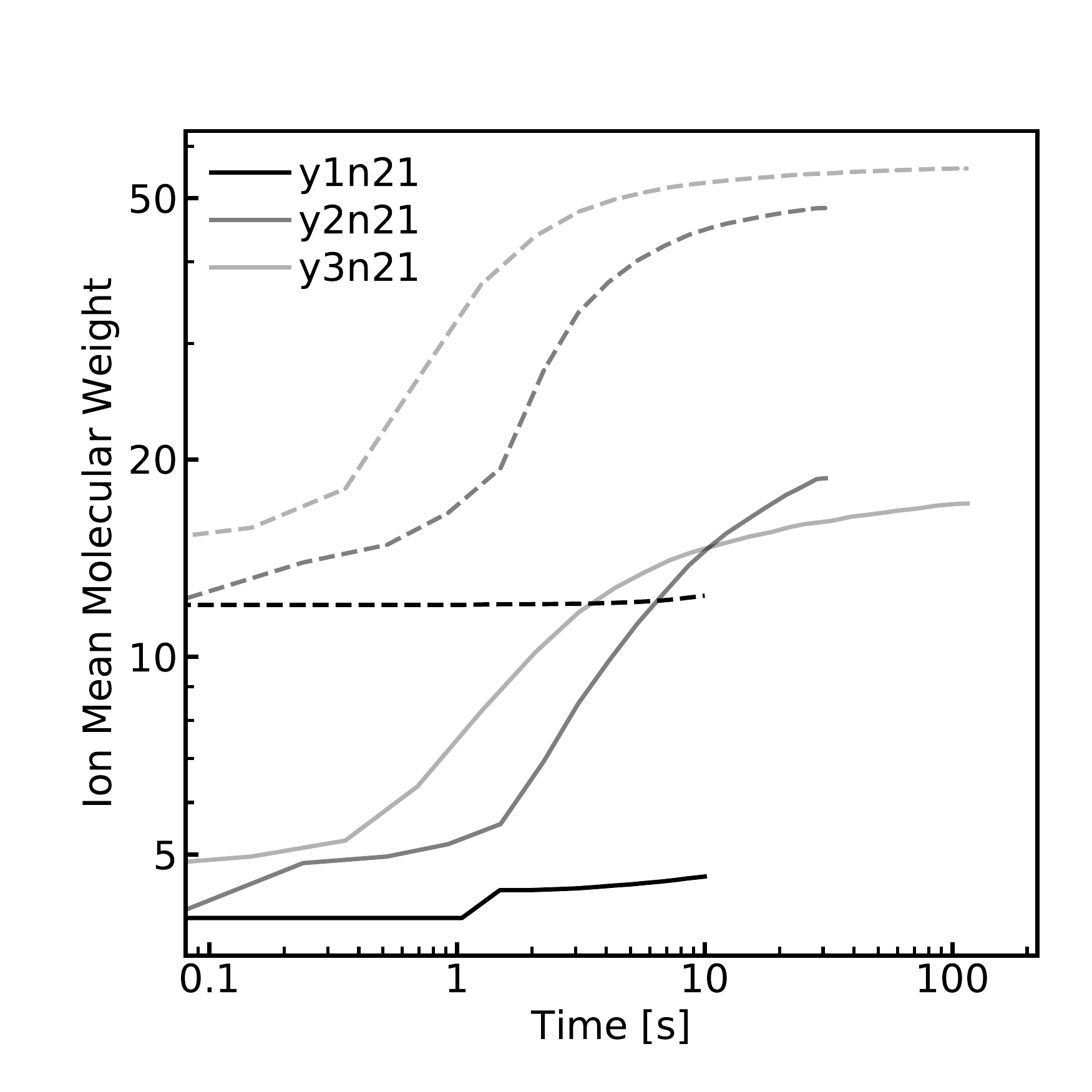}
\caption{Ion mean molecular weight (solid lines) at the photosphere as a function of time.  The dashed lines show the result with $^4$He excluded from the sum over ions.}
\label{fig:mu_vs_t}
\end{figure}

In Figure~\ref{fig:mu_vs_t} we show a related result: the ion mean molecular weight $\mu_{\rm ion}=(\sum_{i}X_i/A_i)^{-1}$  at the photosphere as a function of time, where $X_i$ and $A_i$ are the mass fraction and mass number of element $i$. For reference, a mixture consisting of $50\%$ $^4$He and $50\%$ $^{56}$Ni has $\mu_{\rm ion}=7.5$. For the y2n21 and y3n21 models, it takes a few seconds for $\mu_{\rm ion}$ to increase above 7.5, agreeing well with the $t_{\rm ash}$ estimate.  Note that for $t>10\trm{ s}$, the y3n21 model has a smaller $\mu_{\rm ion}$ than the y2n21 model. This is because a larger amount of $^{4}$He is left unburned in y3n21 than in y2n21.\footnote{\citet{Hashimoto:83} showed that for He burning at constant pressure, the mass fraction of unburned $^{4}$He {\it increases} with increasing pressure for $P\ga 10^{22}\trm{ erg cm}^{-3}$ (see their Figure 10). The pressure at the base of the burning layer $P_b\simeq g y_b\approx 10^{23} y_{b,9}\trm{ erg cm}^{-3}$ is nearly constant during a burst.} To illustrate this effect, the dashed lines in Figure~\ref{fig:mu_vs_t} show the ion mean molecular weight excluding $^4$He. The value for the y3n21 model approaches 56 since it is dominated by $^{56}$Ni while the y2n21 model approaches 48 since it is dominated by $^{48}$Cr. 

\section{Observational Signatures}
\label{sec:observational_signatures}

By evaluating quantities at the photosphere of our wind models, we  calculate theoretical burst light curves and spectroscopy (Section~\ref{sec:spectra_lightcurve}) and then compare our results to observed PRE bursts (Section~\ref{sec:comp_pre}).

\subsection{Burst Spectroscopy and Light Curve}
\label{sec:spectra_lightcurve}

 \begin{figure}
\includegraphics[trim=-0.4cm 0 0 -0.8cm, width=0.45\textwidth]{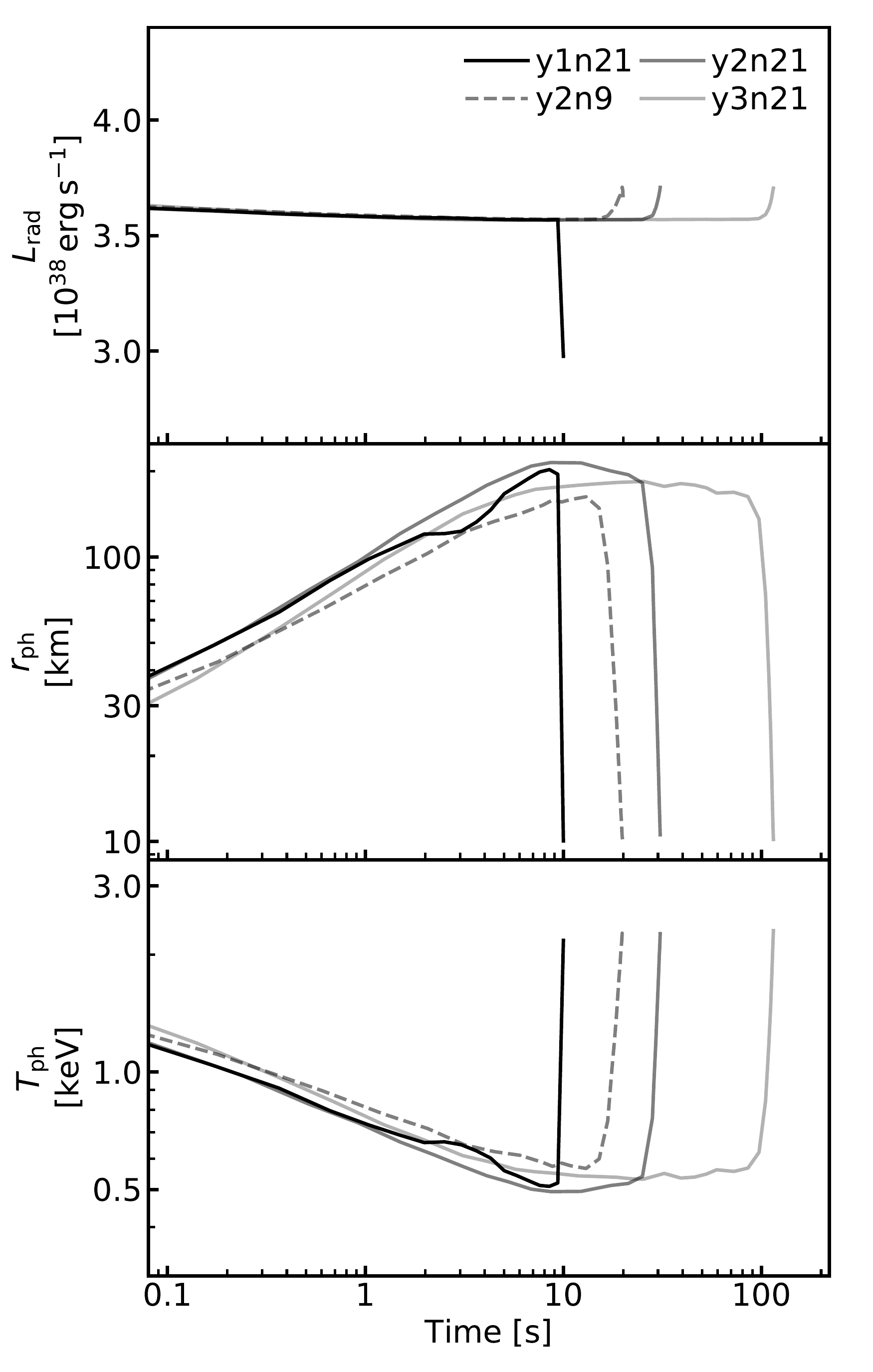}
\caption{Evolution of the bolometric luminosity  $\Lrad$ at the photosphere, the photospheric radius $r_{\rm ph}$, and the photospheric temperature $T_{\rm ph}$, for four different models.}
\label{fig:wind_hist}
\end{figure}

In Figure~\ref{fig:wind_hist} we show the evolution of the bolometric luminosity, photospheric radius $r_{\rm ph}$, and photospheric temperature $T_{\rm ph}$, for different ignition models (these are related by $\Lrad = 4\pi r_{\rm ph}^2\sigma T_{\rm ph}^4$).  Depending on the ignition depth, the PRE phase lasts from $\simeq 5\trm{ s}$ to $\simeq 100\trm{ s}$  (see Equation \ref{eq:tw}), during which the bolometric luminosity is nearly constant at $\Lrad \simeq L_{\rm Edd}$.  
Initially, all the models follow similar tracks: from $t=0\trm{ s}$ to $t\approx 3\trm{ s}$, the photosphere expands from $10\trm{ km}$ to $\ga 100\trm{ km}$ and the temperature $T_{\rm ph}\propto r_{\rm ph}^{-1/2}$ drops from about $2$ to $0.5 \trm{ keV}$.
Due to the larger nuclear energy release, the deeper ignition models expand outward for longer and reach slightly larger radii ($150-200\trm{ km}$).  Furthermore, unlike model y1n21, which shows a fairly abrupt contraction after reaching maximum expansion, models y2n21 and y3n21 have a long, approximately steady phase during which  $r_{\rm ph}$ remains near its maximum for $\simeq 20\trm{ s}$ and $\simeq 90\trm{ s}$, respectively.  Comparing models y2n9 and y2n21, we see that the latter expands faster and reaches a larger maximum $r_{\rm ph}$ because it has a more complete reaction network and thus a larger energy release.

 \begin{figure}
\includegraphics[trim=-0.4cm 0 0 0.5cm,width=0.45\textwidth]{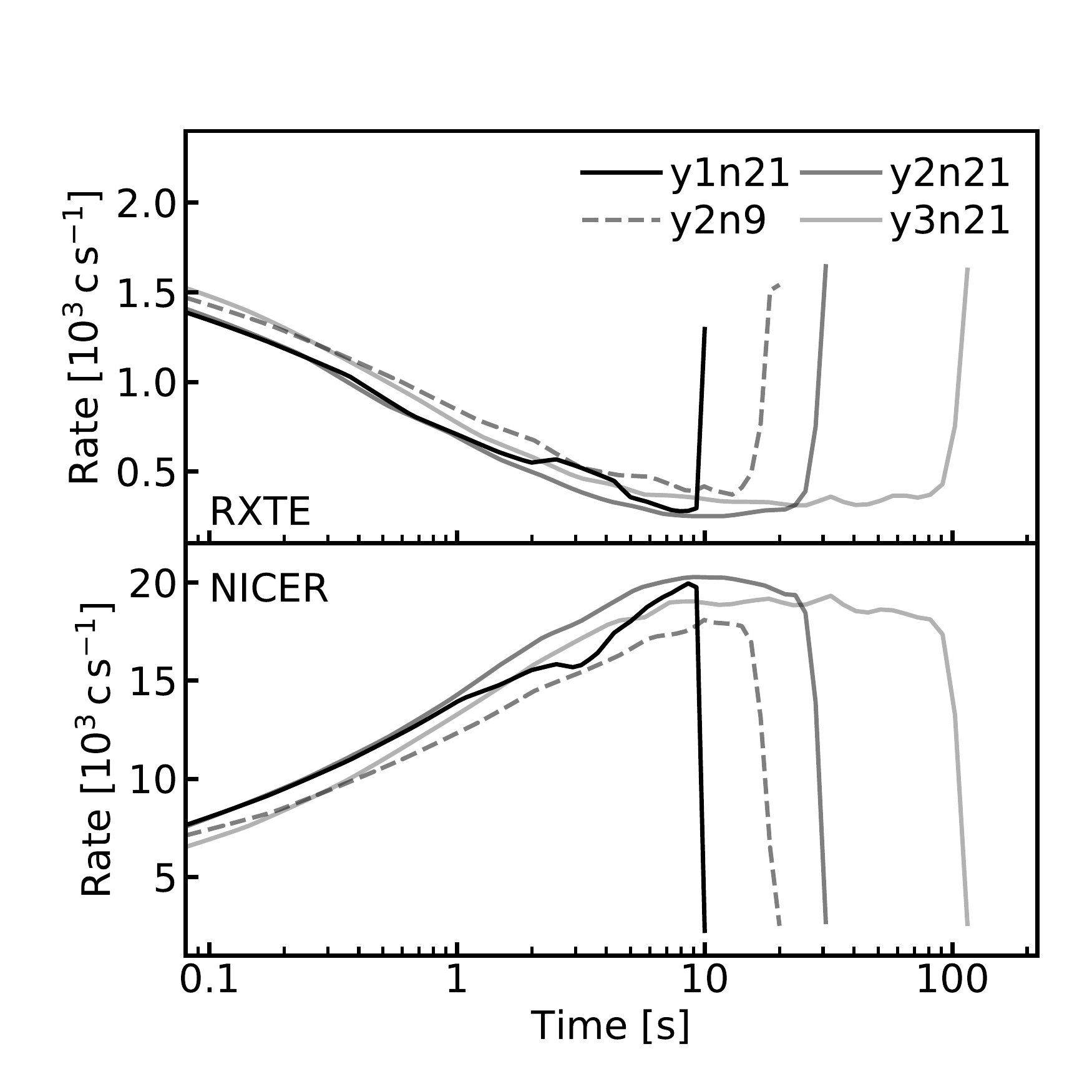}
\caption{Photon count rate as a function of time assuming a source at a distance of 10\,kpc. The upper panel assumes a {\it RXTE}-PCA-like detector, and the lower panel assumes a {\it NICER}-like detector. }
\label{fig:xray_flux}
\end{figure}

In Figure~\ref{fig:xray_flux} we show the approximate count rate that {\it RXTE}-PCA and {\it NICER} would detect for a source located at a distance of $10\trm{ kpc}$.  We assume a blackbody spectrum and use our calculation of $\Lrad[r_{\rm ph}(t)]$ and $T_{\rm ph}(t)$ to estimate the count rate integrated over the effective area of the detector. For {\it RXTE}-PCA, we take the effective area from \cite{Jahoda:06} and assume that two out of the five proportional counting units are working, as was  typical during its operations \citepalias{intZand:10}.
For {\it NICER} we adopt the effective area given on the mission website\footnote{\url{https://heasarc.gsfc.nasa.gov/docs/nicer/}}. The effective collecting area of {\it RXTE}-PCA decreases significantly below 2\,keV and therefore its count rate drops during a PRE, as the spectrum shifts to lower $T_{\rm ph}$. By contrast, the effective collecting area of {\it NICER} remains large down to 0.3\,keV and its count rate actually increases during a PRE.  Moreover, since $T_{\rm ph}\la 1\trm{ keV}$ throughout the PRE, the {\it NICER} count rate is always significantly larger that {\it RXTE}'s (note the different scales in Figure~\ref{fig:xray_flux}).  Both of these effects have been reported in the {\it NICER} observation of 4U 1820-30 \citep{Keek:18}.

In Section~\ref{sec:ejection_of_heavy_elements} we showed that the wind ejects heavy elements synthesized during the burst.  These ejected ashes include $^{48}$Cr, $^{44}$Ti, and $^{56}$Ni at mass fractions $X\ga 0.1$.  \citetalias{Weinberg:06} showed that the expanded photosphere is sufficiently cool that heavy elements such as these will bind with electrons and imprint significant photoionization edges on the burst spectra.  However, they assumed that the wind base is located at $y_{\rm wb}= 0.01 y_b$ whereas our calculations explicitly determine $y_{\rm wb}(t)$ and show that $y_{\rm wb}\approx \trm{few}\times 10^{-3}y_b$ at its maximum (see Figure~\ref{fig:eta_vs_yb}).  \citetalias{Weinberg:06} estimated that during the PRE the edges should have equivalent widths $\trm{EW}\sim 0.1\trm{ keV}$.  Using their approach for estimating the edge strengths and the abundances from our wind calculation, we also find $\trm{EW}\sim 0.1\trm{ keV}$ during the PRE for bursts with $y_b\ga 5\times10^8\trm{ g cm}^{-2}$.

\subsection{Comparison to Observed PRE Bursts}
\label{sec:comp_pre}

Our $r_{\rm ph}(t)$ results are broadly consistent with observations of  PRE bursts,  although there are some notable differences.  According to \citetalias{intZand:10}, the vast majority ($\ga 99\%$) of  PRE bursts have photospheres that do not expand beyond $10^3\trm{ km}$ (the exceptions are superexpansion bursts, which we discuss below).  This is consistent with our result that $r_{\rm ph}\sim100\trm{ km}$ nearly independent of ignition depth. There are weak PREs with maximum $r_{\rm ph}$ that are only a factor of a few larger than $R$ \citep{Galloway:08}.  These PREs may be weaker because they are igniting in mixed H/He layers, and thus by assuming a pure He layer our simulations do not capture this population.

One of the most well sampled measurements of $r_{\rm ph}(t)$ is from the recent burst detected with {\it NICER} from 4U 1820-30 \citep{Keek:18}. This source is an ultracompact X-ray binary (UCXB) that is thought to be accreting  He-rich material \citep{Cumming:03}.  As Figure~\ref{fig:xray_flux} illustrates, {\it NICER}  is ideally suited to follow the entire PRE phase of bursts due its sub-keV sensitivity.  \citet{Keek:18} found that the entire PRE phase lasts $\simeq 3\trm{ s}$ and reaches a maximum expansion radius $r_{\rm ph}=190\pm 10\trm{ km}$.  These are consistent with the duration and expansion radius of our y1n21 model ($y_b=5\times10^8\trm{ g cm}^{-2}$).

On the other hand, the temporal variation of the count rate and $r_{\rm ph}$ of the 4U 1820-30 burst look somewhat different from our y1n21 model (compare Figure 3 in \citealt{Keek:18} and our Figs.~\ref{fig:wind_hist} and~\ref{fig:xray_flux}).  The observed expansion timescale at the start of the PRE is only $\simeq 0.1\trm{ s}$ compared to our $\simeq 1\trm{ s}$.  Also,  after reaching maximum expansion, the observed photosphere falls back to the NS surface somewhat more slowly than ours (over $\simeq 3\trm{ s}$ compared to $\simeq 1\trm{ s}$).  A possible explanation for why our expansion is too slow and  contraction is too fast is that we ignore general relativistic effects. \citet{Paczynski:86} showed that at small $\Mdot$, relativistic models predict a larger $r_{\rm ph}$ than Newtonian models  (see their Figures 10 and 11).  For example, 
 at $\Mdot \simeq 4\times10^{17}\trm{ g s}^{-1}$ we find $r_{\rm ph}\simeq35\ {\rm km}$ (see Figure~\ref{fig:wind_pf_y2n21} at $t=0.07\trm{ s}$) whereas  \citet{Paczynski:86} find $r_{\rm ph}\simeq100\ {\rm km}$.  (The difference is much smaller at $\Mdot \ga 10^{18}\trm{ g s}^{-1}$ and thus our maximum $r_{\rm ph}$ should be reasonably accurate).  As a result, our simulations probably underestimate $r_{\rm ph}$ at the small $\Mdot$ that applies near the start and end of the PRE, which would mean we underestimate (overestimate) the rate of expansion (contraction).

Superexpansion bursts ($r_{\rm ph}>10^3\trm{ km}$) provide another interesting point of comparison.  According to \citetalias{intZand:10}, there have been 32 superexpansion bursts  detected from 8 sources (of these, 22 were from 4U 1722-30).  Of the superexpansion bursts that have been identified with an object (7 out of 8), all are from candidate UCXBs. The neutron star in an UCXB accretes hydrogen-deficient fuel and the bursts tend to be longer (several tens of minutes rather than seconds, i.e., intermediate duration bursts; \citealt{intZand:05, Cumming:06}).  In two superexpansion bursts observed with RXTE, \citetalias{intZand:10} detected strong absorption edges.   The edge energies and depths are consistent with large abundances of iron-peak elements and support our finding that the wind can eject significant amounts of heavy-element ashes. 

Superexpansion bursts always show two distinct phases: a superexpansion phase during which $r_{\rm ph} \ga 10^3\trm{ km}$, followed by a moderate expansion phase during which $r_{\rm ph}\sim 30-50\trm{ km}$ and $\Lrad \simeq \Ledd$ \citepalias{intZand:10}.  Interestingly, the duration of the superexpansion phase is always a few seconds, independent of the ignition depth $y_b$.  By contrast, the duration of the moderate expansion phase ranges from short ($\approx 10-100\trm{ s}$) for $y_b\sim 10^9\trm{ g cm}^{-2}$ to intermediate ($\ga 10^3\trm{ s}$) for $y_b\sim 10^{10}\trm{ g cm}^{-2}$. \citetalias{intZand:10} speculated that the superexpansion phase always lasts a few seconds because it corresponds to a transient stage in the wind's development.  In this stage, they argue, a shell of initially opaque material is ejected to large radii by the sudden onset of super-Eddington flux deep below the photosphere.  Within a few seconds the expanding shell reaches such large radii ($>10^3\trm{ km}$) that it becomes optically thin and the observer suddenly sees the underlying  photosphere of the already formed steady-state wind, which is located at $r_{\rm ph}\sim30-50\trm{ km}$. According to this picture, this marks the onset of the moderate expansion phase, whose duration equals that of the steady-state wind and thus correlates with $y_b$ (see Equation \ref{eq:tw}).

We do not, however, see evidence of a shell ejection in our simulations.  This might be because our treatment of radiative transfer in the low optical depth regions ($\tau \la 3$) is too simplistic, because we ignore relativistic effects, or because of unaccounted for dynamics during the transition from the hydrostatic rise to the hydrodynamic wind  (e.g., \citealt{intZand:14} reported two bursts with exceptionally short precursors, which they argue may indicate a detonation initiated by the rapid onset of the $^{12}\trm{C}(p,\gamma)^{13}\trm{N}(\alpha,p)^{16}\trm{O}$ reaction sequence).   However, there is a feature in our simulations that suggests an alternative explanation for why the superexpansion phase always lasts a few seconds, regardless of $y_b$.  As we described in Section~\ref{sec:ejection_of_heavy_elements}, the timescale $t_{\rm ash}$ to expose heavy elements in the wind is also a few seconds, and plateaus at $t_{\rm ash}\simeq 1\trm{ s}$ for $y_b\gtrsim 3\times10^9\ {\rm g\,cm^{-2}}$.  It suggests that the transition from superexpansion to moderate expansion after $\approx 1\trm{ s}$  might be due to the wind's composition changing from light to heavy elements (and not due to shell ejection). As discussed in \citetalias{intZand:10}, the heavy elements will only be partially ionized and the radiative force  acting on the bound electrons above the photosphere will be $\ga 100$ times larger than the force acting on the free electrons.  Line-driving in the outer parts of the wind may therefore boost the outflow velocity.  By mass conservation (Eq.~\ref{eq:mass_cons}), this would decrease the overlying density and bring the photosphere inward to smaller radii.  This could explain why the photosphere transitions from very large $r_{\rm ph}$ during the superexpansion (when the ejecta are mostly light elements) to smaller values of $r_{\rm ph}\sim 30-50\trm{ km}$ during the moderate expansion, and why the the timescale for such transitions is always a few seconds. Evaluating this in detail likely requires accounting for line-driving in the wind.

\section{Summary and Conclusions}
\label{sec:summary}

We presented spherically symmetric \texttt{MESA} models of PRE bursts, starting from the hydrostatic burst rise through the hydrodynamic wind phase.  We used both a  9-isotope and a 21-isotope reaction network to follow the burning of pure He ignition layers with base column depths $y_b=3\times10^8 - 5\times10^9\trm{ g cm}^{-2}$, corresponding to that of short through intermediate duration PRE bursts.  Convection during the burst rise mixes the ashes of nuclear burning out to $y\la 10^4\trm{ g cm}^{-2}$ for $y_b\ga10^9\trm{ g cm}^{-2}$. As the atmosphere heats up, $\Lrad$ increases and eventually exceeds $\Ledd$, resulting in a radiation-driven wind.   Initially, the wind base is located at 
small column depths but it moves inward to $y_{\rm wb}\ga10^6\trm{ g cm}^{-2}$ in a few seconds. As a result, the wind initially ejects mostly light elements ($^4$He and $^{12}$C) but after $\approx 1\trm{ s}$ it begins to eject heavy elements, which  for $y_b=1.5(5)\times10^9\trm{ g cm}^{-2}$ consists mostly of $^{48}$Cr ($^{56}$Ni).

The wind duration $t_{\rm w}$ increases almost linearly with $y_b$, lasting from a few seconds to more than $100\trm{ s}$ for the considered range of $y_b$.  All $y_b$ show similar wind evolution during the first few seconds: a mass-loss rate that increases to a maximum $\Mdot \simeq 1-2\times10^{18}\trm{ g s}^{-1}$, a photospheric radius that expands out to $r_{\rm ph}\simeq 100-200\trm{ km}$, and a photospheric temperature that  decreases to $\la 0.5\trm{ keV}$.  After the first few seconds, the wind either abruptly dies down (small $y_b$) or it blows steadily for the next $\approx 100\trm{ s}$ (large $y_b$). We found that  the wind ejects $\approx 0.2\%$ of the total accreted mass, nearly independent of $y_b$, which corresponds to $\approx 30\%$ of the nuclear energy release being used to unbind matter from the NS surface.  

Based on the calculated wind composition, we estimated that the ejected heavy elements should imprint photoionization edges on the burst spectra with equivalent widths of $\sim 0.1\trm{ keV}$ for bursts with $y_b\ga 10^9\trm{ g cm}^{-2}$. This supports the evidence of heavy-element absorption features detected in some PRE bursts  (\citetalias{intZand:10}, \citealt{Barriere:15, Iwai:17, Kajava:17}) and encourages efforts to catch strong PRE bursts with high-spectral-resolution telescopes such as {\it Chandra} and {\it XMM-Newton}. 

We showed that our results are broadly consistent with various aspects of observed PRE bursts.  In particular, many PRE bursts show maximum photospheric radii $r_{\rm ph}\sim 100\trm{ km}$, photospheric expansion velocities $v_{\rm ph}\sim 100\trm{ km s}^{-1}$ during the start of the PRE, and PRE durations $t_{\rm w}\sim 1-100\trm{ s}$.  However, we also described some notable differences between our models and observed PRE bursts, which we argued might be because we did not account for general relativistic effects and neglected possible line-driving of heavy elements.  Models that solve the relativistic, time-dependent wind equations and adopt a more sophisticated treatment of radiative transfer are needed.  This includes relaxing the assumption of LTE and the diffusion approximation in the outer parts of the wind and using composition-dependent opacities that account for bound-free and bound-bound transitions, as well as Compton scattering. Such improvements would allow for a more complete understanding of PRE bursts and might help inform PRE-based measurements of NS radii.



\acknowledgments

We thank Deepto Chakrabarty for useful conversations and for providing input on the effective collecting area of  RXTE-PCA and NICER. We are also grateful to Jean in't Zand and the referee for valuable comments on the manuscript.





\appendix

\section{Details of the \texttt{MESA} setup}\label{app:mesa}

Our simulations start from the \texttt{NS\_envelope} model provided in the \texttt{MESA} test suite.  We set  $M=1.4\,M_\odot$ and $R=10\,{\rm km}$ at the inner boundary.  The module provides a thin NS envelope of pure $^{56}$Fe, on top of which we accrete an extra layer of $^{56}$Cr with a column depth of $10^9\trm{ g cm}^{-2}$ (see footnote \ref{fn:dredgeup}). We accrete pure $^4$He onto this envelope until $^4$He ignition using commands similar to the \texttt{ns\_he} model provided in the \texttt{MESA} test suite. 

The column depth of the ignition base, $y_b$, depends on the pre-burst flux $F_{\rm pre}$  \citep{Bildsten:98, Cumming:00}.  For pure helium accretion, $F_{\rm pre}=Q_{\rm crust} \dot{M}_{\rm acc}/4\pi R^2$,  where $\dot{M}_{\rm acc}$ is the accretion rate and $Q_{\rm crust}$ is the energy release per nucleon in the crust~\citep{Brown:98, Cumming:03, Cumming:06}. In \texttt{MESA}, this flux can be controlled by a combination of the mass accretion rate and the core luminosity using the commands ``\texttt{mass\_change}''  and ``\texttt{relax\_initial\_L\_center},'' respectively~\citep{Paxton:11}.
To determine the value of $L_c$, we adopt the following procedure.  First, we 
perform a side calculation in which we evaluate the ignition condition
$d\epsilon_{3\alpha}/dT = d\epsilon_{\rm cool}/dT$,
where $\epsilon_{3\alpha}$ is the triple-alpha energy generation rate and $\epsilon_{\rm cool}=a c T^4/3\kappa y^2$ is the radiative cooling rate.  Here we assume that the thermal profile of the pre-burst atmosphere is given by $acT^4\simeq 3\kappa y F_{\rm pre}$, with $F_{\rm pre}=Q_{\rm crust} \dot{m}_{\rm acc}$.   We determine $Q_{\rm crust}$ by requiring that ignition occur at $y_b=5\times10^{8}\ {\rm g\ cm^{-2}}$ when $\dot{M}_{\rm acc}=3\times10^{-9} M_\odot\ {\rm yr}^{-1}$.\footnote{ These are typical ignition parameters (see, e.g., Table 1 of \citetalias{Weinberg:06}); we find very similar wind simulation results when we instead require $y_b=3\times10^8\ {\rm g\ cm^{-2}}$ at $\dot{M}_{\rm acc}=3\times10^{-9} M_\odot\ {\rm yr}^{-1}$.}  With $Q_{\rm crust}$ fixed at this value, we obtain an ignition curve $y_b(\dot{M}_{\rm acc})$.  We then determine the value of  $L_c$ by requiring that the \texttt{MESA} model ignite at a depth that matches the ignition curve $y_b(\dot{M}_{\rm acc})$.  For the $\{y1, y2, y3\}$ models, we find $(\dot{M}_{\rm acc}, L_c)$ equals $\{(3, 2.45), (0.5, 1.36), (0.08, 0.85)\}$,  in units of $(10^{-9}\ M_\odot\ {\rm yr}^{-1}, 10^{34}\ {\rm erg\ s^{-1}})$.

Once the base becomes convective, we use the following inlist file to simulate the hydrostatic burst rise.

\begin{tabular}{ll}
\toprule
\toprule
\begin{minipage}{3in}
\begin{verbatim}
&star_job
  change_initial_net = .true.
  new_net_name = `approx21.net'
  kappa_file_prefix = `gs98'
  relax_initial_tau_factor = .true.
  relax_to_this_tau_factor = 100d0
  dlogtau_factor = .1
  change_v_flag = .true.
  change_initial_v_flag = .true.
  new_v_flag = .true.
&controls    
  max_timestep = .5d-2
  use_GR_factors = .false.
  varcontrol_target = .75d-4
  which_atm_option = `grey_and_kap' 
  Pextra_factor = 2
  accrete_same_as_surface = .false. 
  accrete_given_mass_fractions = .true. 
  num_accretion_species = 1
  accretion_species_xa(1) = 1
  accretion_species_id(1) = `he4'
  mass_change = 5d-10
  mixing_length_alpha = 1.5
  MLT_option = `Henyey'    
  okay_to_reduce_gradT_excess = .false.
  use_Ledoux_criterion = .false.
\end{verbatim}
\end{minipage}
\\
\bottomrule
\end{tabular}
\\
\\

We use the `\texttt{approx21.net}' network for the n21 models and the `\texttt{basic\_plus\_fe56.net}' for the n9 models. For simplicity, we neglect the composition gradient's contribution to convection (\texttt{use\_Ledoux\_criterion = .false}), since we assume pure He accretion. Such a simplification would not be appropriate for mixed H/He accretion~\citepalias{Weinberg:06}. The choice of \texttt{mixing\_length\_alpha} is motivated by the value given in the test suite \texttt{ns\_he}, although we find that the results are not particularly sensitive to this choice.  In order to be consistent with the subsequent hydrodynamic simulations, we turn off the \texttt{MLT++} option by setting \texttt{okay\_to\_reduce\_gradT\_excess=.false.}; using it in the hydrodynamics calculation would have suppressed the generation of a wind \citep{Paxton:13, Quataert:16}. 

We edited the \texttt{extras\_check\_model} function in \texttt{run\_star\_extras.f} in order to stop the hydrostatic simulation when $\Lrad$ first exceeds $\Ledd$ at the top boundary.  We pass the final profile of the hydrostatic simulation to the hydrodynamic simulation, which uses the following inlist file.
\begin{tabular}{ll}
\toprule
\toprule
\begin{minipage}{3in}
\begin{verbatim}
&star_job
  relax_initial_tau_factor = .true.
  relax_tau_factor = .true.
  relax_to_this_tau_factor = 1
  dlogtau_factor = .1    
  set_initial_dt = .true.
  seconds_for_initial_dt = 1d-4    
  remove_surface_by_density = 1d-14
  repeat_remove_surface_for_each_step =.true.    
&controls  
  varcontrol_target = 2d-5    
  MLT_option = `none'
  Dutch_scaling_factor = 0.0
  Dutch_wind_lowT_scheme = `de Jager'
  Hot_wind_scheme = `Dutch'    
  which_atm_option = `grey_and_kap' 
  Pextra_factor = 3
  use_compression_outer_BC = .true.
  use_zero_dLdm_outer_BC = .true.    
  shock_spread_linear = 0.0
  shock_spread_quadratic = 1d-2    
  use_ODE_var_eqn_pairing = .true.
  use_dPrad_dm_form_of_T_gradient_eqn = .true.
  use_dvdt_form_of_momentum_eqn = .true.
  use_ODE_form_of_density_eqn = .true.    
  okay_to_remesh = .true.
  min_dq = 1d-12 
  max_center_cell_dq = 5e-6
  max_allowed_nz = 14000
  max_surface_cell_dq = 1d-12 
  P_function_weight = 40
  log_tau_function_weight = 22
  log_kap_function_weight = 20
  logQ_min_limit = -18d0   
  newton_iterations_limit = 7
  iter_for_resid_tol2 = 4
  tol_residual_norm1 = 1d-8
  tol_max_residual1 = 1d-7
  fe_core_infall_limit = 1d99
  tiny_corr_coeff_limit = 999999
  newton_itermin_until_reduce_min_corr_coeff 
                                     = 999999
\end{verbatim}
\end{minipage}
\\
\bottomrule
\end{tabular}
\\
\\
The values controlling the numerical mesh and function weights are for the y2n21 model; in order to achieve convergent solutions, some models require slightly different settings (e.g., different values for \texttt{varcontrol\_target}, \texttt{log\_tau\_function\_weight}, \texttt{min\_dq},  and \texttt{max\_surface\_cell\_dq}). Typically, each profile has approximately 1000 zones in our simulations.

\bibliography{ref}

\end{document}